\documentclass{bio}
\usepackage{natbib}
\newcommand{\given}{\mathbin{\vert}}
\newcommand{\prob}[1]{\mathbb{P}\left(#1\right)}
\newcommand{\expected}[1] { \operatorname{\mathbb{E}} \left[ #1 \right]}
\newcommand{\indic}[1]{ \mathbb{I}{\left[#1\right]}}
\newcommand{\indep}{\rotatebox[origin=c]{90}{$\models$}}

\begin{document}

% Title of paper
% linear mixed models to compare dynamic treatment regimens on a longitudinal outcome
% Mixed effects
% in sequentially randomized trial
\title{Linear Mixed Models for Comparing Dynamic Treatment Regimens on a Longitudinal Outcome in Sequentially Randomized Trials}

% List of authors, with corresponding author marked by asterisk
\author{BROOK LUERS$^\ast$\\
% Author addresses
\textit{Department of Statistics, 
University of Michigan,
Ann Arbor, MI}\\[2pt]
% E-mail address for correspondence
{luers@umich.edu}\\
MIN QIAN\\
% Author addresses
\textit{Department of Biostatistics, Mailman School of Public Health,
 Columbia University, New York, NY}\\
%{mq2158@columbia.edu}\\
INBAL NAHUM-SHANI\\
% Author addresses
\textit{Survey Research Center, Institute for Social Research, University of Michigan, Ann Arbor, MI }
%{inbal@umich.edu}\\
CONNIE KASARI\\
\textit{Department of Education and Semel Institute for Neuroscience,
 University of California, Los Angeles, CA}\\
% {kasari@gseis.ucla.edu}\\
 DANIEL ALMIRALL\\
 \textit{Survey Research Center, Institute for Social Research and Department of Statistics, 
 University of Michigan, Ann Arbor, MI}\\
% {dalmiral@umich.edu}
}

% Running headers of paper:
\markboth%
% First field is the short list of authors
{B. Luers and others}
% Second field is the short title of the paper
{Linear Mixed Models for Sequentially Randomized Trials}

\maketitle

% Add a footnote for the corresponding author if one has been
% identified in the author list
\footnotetext{To whom correspondence should be addressed.}

\begin{abstract}
{A dynamic treatment regimen (DTR) is a pre-specified sequence of decision rules 
which maps baseline or time-varying measurements on an individual to a 
recommended intervention or set of interventions. 
Sequential multiple assignment randomized trials (SMARTs) 
represent an important data collection tool for informing the construction of effective DTRs. 
% Most SMARTs have multiple DTRs embedded within the trial. 
A common primary aim in a SMART is the marginal mean comparison between 
two or more of the DTRs embedded in the trial.
This manuscript develops a mixed effects modeling and 
estimation approach for these primary aim comparisons 
based on a continuous, longitudinal outcome. 
The method is illustrated using data from a SMART in autism research.}
{Adaptive interventions; 
Sequential multiple assignment randomized trials; 
SMART; causal inference; mixed effects modeling}
\end{abstract}

\section{Introduction}
\label{sec:intro}
A dynamic treatment regimen (DTR) is a pre-specified sequence 
of decision rules which map baseline and time-varying measurements
on an individual to a recommended set of interventions
\citep{
%laber2014dynamic,
chakraborty-moodie_DTR-Book:2013,
orellana-rot-robins_dynamicMSM_partI:2010,
hernan/optimaldtr-via-iptw:2006,
murphy/laan/robins/cpprg:01}.
DTRs are designed to 
assist clinicians with ongoing care decisions
based on disease progress, treatment history, 
%engagement or adherence history, 
and other
information collected during the course of treatment. 
DTRs are also known as adaptive treatment strategies 
\citep{kosorok2015adaptive,
%Wallace2014,
murphy-lynch-DAD:2007} 
or adaptive interventions 
\citep{
%almirall2016adaptive,
almirall_IntroToSMART:2014,
nahum_SMARTprimaryPsychMeth:2012}.

A sequential,
multiple assignment randomized trial (SMART)
is a multi-stage trial design
specifically created for comparing or constructing DTRs 
\citep{%wallace2016smart,
%kosorok2015adaptive,
lavori2014introduction,
%chakraborty-moodie_DTR-Book:2013,
lei_SMART:2012,
murphy_SMARTsim:2005}. 
%The SMART design consists in a 
%sequence of randomizations corresponding to the
%sequence of decision rules in a DTR. 
Study participants in a SMART may experience multiple
randomizations.
These randomizations occur at decision points 
for which there is a question about which treatment to provide. 
By the end of the trial, 
specific groups of study participants will have been subject
to the sequence of treatment decisions corresponding
to at least one of a pre-specified set of DTRs. 
SMARTs enable causal comparisons among 
these ``embedded'' DTRs.

In this article we focus on scientific questions
which involve comparing the embedded DTRs 
in a SMART based on 
the mean of a continuous, longitudinal outcome. 
 Often this is a primary scientific aim in a SMART 
 \citep{seewald2018-repeatedmeasures-power-compareDTRs}.
One way of answering these questions involves 
directly specifying a model for the marginal mean
of the longitudinal outcome under each DTR 
 and estimating the parameters in that model using 
 weighted estimating equations
\citep{lu:2016,
 seewald2018-repeatedmeasures-power-compareDTRs}.
% nahum-shani2018-tutorial-repeatedmeasures-compareDTRs-SMART
 % almirall2016longitudinal : analysis of autism study, longitudinal effects
Similar methods are available when the longitudinal outcome is binary 
\citep{dziak-binary-2019},
for a survival outcome
\citep{Zhiguo-survival}, 
and for clustered SMARTs
where the embedded DTRs are applied to clusters
of people but outcomes
are measured on individuals within each cluster 
\citep{necamp2017comparing-DTRs-cluster-SMART}.

The purpose of this article is to develop
 linear mixed effects models 
 for primary aim comparisons of 
 the embedded DTRs in a SMART with a continuous,
 longitudinal outcome.
Mixed effects models are a well established tool 
for analyzing longitudinal, clustered, or multilevel 
data in the medical, social, and agricultural sciences
\citep{fitzmaurice2012applied,
raudenbush2002hierarchical,
snijdersmultilevel,
searlevariance,
% verbeke2009linear,
goldstein2011multilevel,
hedeker2006longitudinal}. 
This paper provides a way for researchers to analyze data
from SMARTs using these familiar statistical tools. 
In addition, we see at least three reasons why scientists might prefer 
mixed effects models when analyzing SMARTs.

First, mixed models provide an intuitive, flexible way to model 
within-person correlations among longitudinal outcomes. 
Existing statistical methods for %comparing embedded DTRs in a 
SMARTs 
 with a continuous, longitudinal outcome
\citep{lu:2016, 
seewald2018-repeatedmeasures-power-compareDTRs}
%account for correlation among longitudinal measurements 
involve directly specifying a working model for the marginal
covariance matrix of the repeated measures, as in 
% which is then
%treated as a nuisance parameter. 
generalized estimating equations 
(GEE, \citet{liang1986longitudinal}).
In contrast, our mixed effects model 
indirectly parameterizes the marginal covariance 
using random effects---latent random variables 
which describe subject-specific change over time. 
This specification distinguishes within-subject and between-subject 
variation and provides an intuitive and flexible 
way to model the marginal covariance as 
a function of time and other covariates.

Second, modeling the within-person correlation 
among longitudinal measurements can 
improve statistical efficiency in estimating regression parameters 
\citep[e.g.][Section 4.6]{diggle2002analysis}, 
and mixed models easily parameterize rich covariance functions using few parameters, 
regardless of the number or spacing of measurement occasions 
(\citet[Chapter 8]{fitzmaurice2012applied}, \citet[Chapter 8]{hedeker2006longitudinal}). 

Third, mixed models provide predictions of subject-specific outcome trajectories via 
prediction of the random effects 
(\citet{skrondal2009prediction}, 
\citet[Chapter 4]{hedeker2006longitudinal},
\citet[Chapter 7]{searlevariance}).
While such predictions do not constitute the primary aim of 
comparing embedded DTRs in a SMART, 
they may be useful in understanding the type and magnitude of 
heterogeneity in person-specific change
with respect to the embedded DTRs 
% over time
 and in identifying individuals with unusual response trajectories. 

Throughout this paper we will refer to an example SMART designed to compare
three DTRs for improving spoken language in children with autism. 
Section~\ref{sec:smart} introduces this study design and provides
 a general description of SMARTs and embedded DTRs. 
 Section~\ref{sec:LMM} introduces our proposed mixed model 
 for comparing embedded DTRs in a SMART and 
 Section~\ref{sec:estimation-prototypical} describes
 how we estimate parameters and predict random effects in this model. 
In Section~\ref{sec:sims} we report the results of simulation 
experiments which investigate the operating characteristics of our estimation method, 
and in Section~\ref{sec:application} we illustrate the method using data from the autism SMART
introduced in Section~\ref{sec:smart}. 

\section{Sequential, Multiple-Assignment Randomized Trials}
\label{sec:smart}

Sequential, multiple assignment randomized trials (SMARTs) are multi-stage randomized trial designs
which were developed explicitly for the purpose of building high-quality
DTRs 
\citep{murphy_SMARTsim:2005,
lavori-dawson_biasedRandomize:2000,
dawson2008sequential}. 
Each participant in a SMART may move through multiple stages of treatment, and the
 defining feature of a SMART is that some or all participants are randomized at more 
than one decision point. At each decision point, the purpose of randomization is 
to address a question concerning
the dosage %(duration, frequency or amount), 
intensity, type, or delivery of treatment at
that decision point. 
A common primary aim in a SMART is the marginal mean comparison of two or 
more embedded DTRs on a longitudinal research outcome. 
The following example SMART illustrates these ideas.

\subsection{An Example SMART in Autism}
\label{sec:autism-desc}
\begin{figure}[!p]
\includegraphics[width=\textwidth]{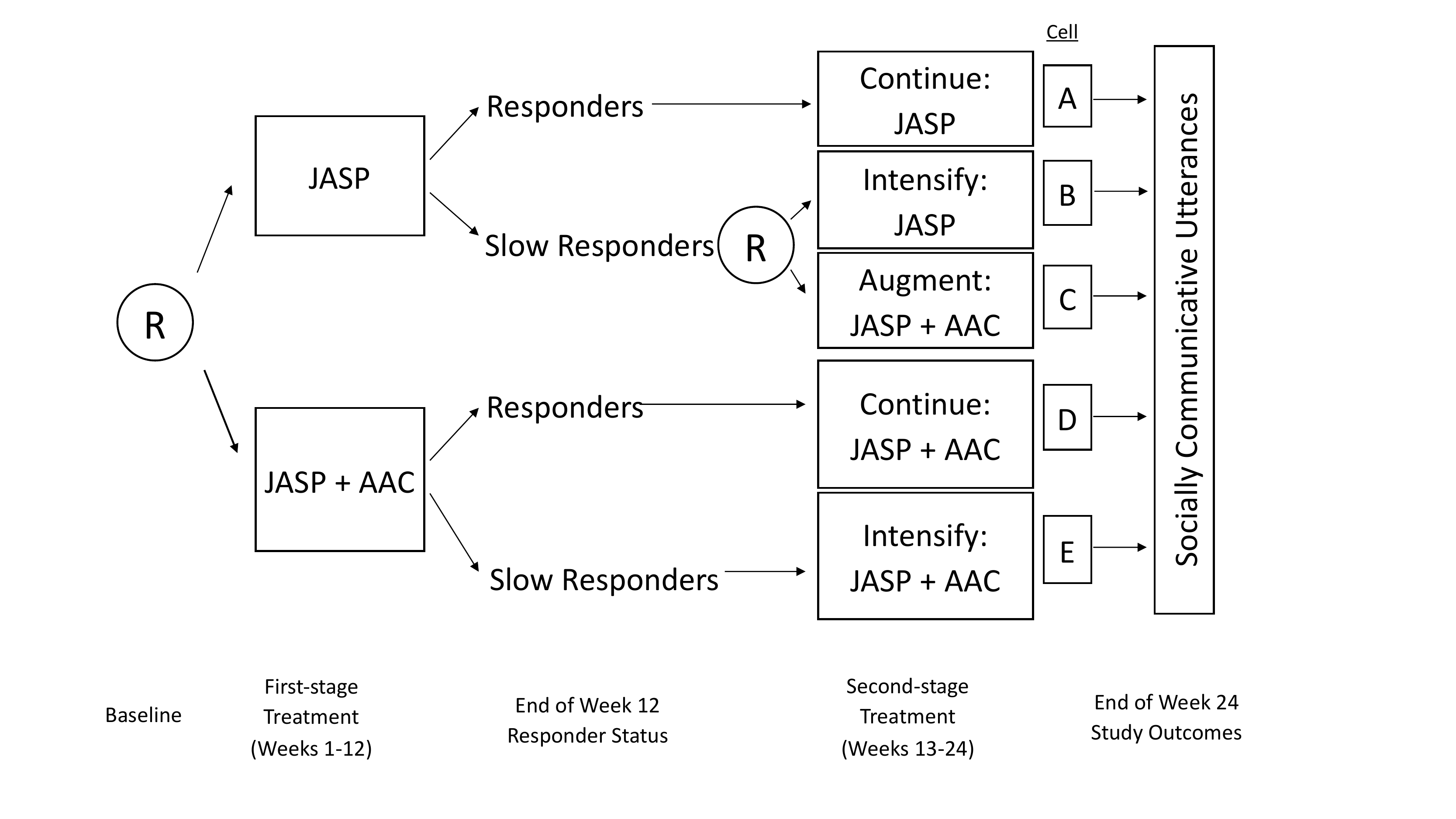}
\caption{Schematic of an example sequential multiple-assignment randomized trial (SMART) for children with ASD who are minimally verbal. JASP stands for joint attention social play intervention; AAC stands for alternative and augmentative communication. The encircled R signifies randomization; randomizations occurred at baseline and at the end of week 12 following identification of responder status. A child was considered a responder if there is a 25\% or greater improvement on 7 or more (out of 14) language measures; otherwise, the child was labeled a slow responder.}
\label{fig:exampleSMART_autism}
\end{figure}

The SMART shown in Figure~\ref{fig:exampleSMART_autism} 
\citep{kasari2014communication} 
involved $N=61$ children, between five and eight years old, 
who had a previous diagnosis of autism spectrum 
disorder and were considered ``minimally verbal'' 
(used fewer than 20 spontaneous different words during a baseline 20-minute language test). 
%All of the children were between five and eight years old and 
%had at least two years of prior treatment 
All eligible children were initially randomized, with equal probability, 
to a behavioral treatment, called JASP, 
or to JASP together with a speech-generating device, 
called AAC (augmentative or alternative communication).
Both of these first-stage treatment arms in the SMART involved twice-weekly
sessions with a trained language therapist. The first-stage JASP+AAC arm required
that the AAC device was used at least 50 percent of the time during these sessions.

At the end of the first treatment stage, which lasted 12 weeks,
 all children were classified as ``responders'' or ``slow responders''. 
Response was defined, prior to the trial, as an improvement of at least 25 percent
on seven or more % out of 14
language measures (e.g. words used per minute) % and mean length of utterances)
by the end of week 12.
%25 percent, or greater improvement from baseline to 
%week 12 on seven or more of 14 language measures 
%(e.g. words used per minute and mean length of utterances).
Children who did not satisfy this criterion were considered slow responders. 

 The second-stage treatments were determined as follows. 
 Responders to the initial intervention were continued on that intervention for an additional 12 weeks. 
 Slow responders to JASP+AAC were offered intensified JASP+AAC, 
 which involved increasing the number of weekly sessions from two to three. 
 Slow responders to JASP were re-randomized, with equal probability,
  to either intensified JASP 
 %(increasing the number of weekly sessions from two to three) 
 or to JASP+AAC. 
% Parents were involved in all aspects of intervention, including in a parent training component. 
 The status of ``responder'' or ``slow responder'' in this SMART is known as 
 an \emph{embedded tailoring variable}, since it is used to restrict subsequent 
 randomizations and is therefore a part of the embedded DTRs. 
% which are described in Section~\ref{sec:dtr}. 
 The primary research outcome in this SMART was the total number of spontaneous socially 
 communicative utterances in a 20-minute language sample, 
 measured by an independent evaluator who was blind to the
 assigned treatment sequence. 
This primary outcome was measured four times:
 prior to the initial randomization (baseline), 
 prior to the second randomization (at week 12), 
 at the end of treatment (week 24) and at a follow-up assessment (week  36).

\subsection{Dynamic Treatment Regimens Embedded in a SMART}
\label{sec:dtr}
A dynamic treatment regimen (DTR) is a sequence of decisions rules that, 
for all individuals in a population of interest, 
guides the provision of treatment at each 
decision point based on information known up to that decision point. 
In the case of the autism SMART, a DTR is a sequence of decision 
rules that guides the first and second treatment decisions for 
both responders and slow responders. 

Specifically, the autism SMART has three DTRs embedded within it.
These are listed in Table~\ref{table:embeddedDTRs}. 
The DTR labeled (AAC, AAC+) starts with JASP and AAC,
continues this treatment for responders 
 and intensifies this treatment for slow responders.
 The other two DTRs
start with JASP only. For slow responders, (JASP, JASP+)
intensifies JASP alone while (JASP, AAC) augments JASP with
AAC. 
Many SMARTs 
use a two-stage design in which only slow responders
are randomized at the start of the second stage
\citep[e.g.]{kidwell-SMARTcancerResearch:2014,
gunlicks2015pilot}.
%This SMART is somewhat atypical in that 
In this SMART, however, second-stage randomization
was restricted based on a combination of first-stage treatment
and response status. 
We use $(a_1,a_2)$ to index the DTRs embedded in the SMART,
where $a_j$ denotes the treatment provided at the $j$th decision point. 
Table~\ref{table:embeddedDTRs} enumerates the values of 
$(a_1,a_2)$ for each DTR in the autism SMART.

\section{Linear Mixed Models for Comparing Embedded DTRs}
\label{sec:LMM}

We aim to develop a linear mixed model for primary aim 
comparisons based on a pre-specified summary of the mean outcome under
each DTR in a SMART.
To do this, we use the 
potential outcomes framework to describe the sequence of
 primary outcome measurements
as a function of the embedded DTRs. 
For simplicity, we focus on two-stage designs. 
With slight changes in notation, the methodology presented here 
may be generalized to more complex SMART designs.
% with,
%for example, more than two treatment stages or multiple randomizations 
%for both responders and slow responders.

\subsection{Potential outcomes and observed data} 
\label{sec:notation}
For each embedded DTR, indexed by $(a_1,a_2)$ where 
$a_1,a_2 \in \left\{-1, +1\right\}$, and for 
the $i$th SMART participant, $i=1,\ldots,N$, let
$Y_{i}(a_1,a_2) = (Y_{it_{i1}}(a_1,a_2), Y_{it_{i2}}(a_1,a_2), \ldots, Y_{it_{in_i}}(a_1,a_2))^\intercal$
denote the vector of $n_i$ time-ordered, potential outcome repeated measures.
The vector 
$Y_i(a_1,a_2)$ is simply the set of longitudinal potential outcomes 
for participant $i$ under DTR $(a_1,a_2)$. 
For example, in the case of the 
autism SMART, each participant has three
potential values of $Y_i(a_1,a_2)$, corresponding to the 
three values for $(a_1,a_2)$ given in Table~\ref{table:embeddedDTRs}. 
Note that in the autism study, $a_2$ 
is undefined for the DTR beginning with JASP+AAC, since that DTR is
 fully characterized by $a_1=-1$ (slow responders to JASP+AAC were not re-randomized). 
%In the autism SMART and in many other common designs, 
%the embedded tailoring variable 
%is a binary summary of the data collected during the first stage
%and often represents early or delayed response to 
%first-stage treatment. 
Let $R_i(a_1) \in \left\{0,1\right\}$ 
be the potential outcome for the binary embedded tailoring variable 
under first-stage treatment $a_1$. 

During the conduct of a SMART, we collect the following 
observed data: 
 $Y_{it_{ij}}$, the observed primary outcome for participant $i$ at time point $t_{ij}$; 
 $R_i$, the $i$th participant's observed binary tailoring variable; 
 $L_i$, a pre-specified vector of baseline covariates collected prior to the first randomization; 
 and $A_{1i}, A_{2i}$, the random treatment assignments 
 in the first and second stage, respectively. 

%Below, we sometimes use $Y_{it}$ in place of $Y_{ij}$ to denote 
%an observed primary outcome collected at generic time point $t$.
 In the autism SMART, the primary outcome was 
 collected for all children at each of $n_i=4$ measurement occasions, 
occurring 12 weeks apart. So in this example we let 
$t_{ij}=t_j \in \left\{0, 12, 24, 36\right\}$ denote the time, in weeks,
since baseline assessment. 
%Thus, the primary outcome in this study 
%(spontaneous socially communicative utterances) 
%was 
%collected at $t_{1}=0$, $t_{2}=12$, $t_{3}=24$ and $t_{4}=36$. 
In the autism SMART, $A_{1i}$ is equal to $1$ or 
$-1$ with equal probability, indicating randomization to either 
JASP or JASP+AAC. Among slow responders to JASP, that is, 
among all subjects with $A_{1i}=1$ and $R_i=0$, $A_{2i}$ 
equals $1$ or $-1$ with equal probability, denoting randomization to receive
either intensified JASP or the AAC device. 
In the autism study, $A_{2i}$ is not defined for 
responders and participants randomized to $A_{1i}=-1$.

\subsection{The model}
\label{sec:themodel}
% START HERE OCTOBER 12
For the $i$th participant 
%at the $j$th measurement occasion 
and for a fixed DTR $(a_1,a_2)$, 
consider the following linear mixed effects model:
\begin{equation}
Y_{i}(a_1,a_2) =  X_{i}(a_1,a_2)\beta + Z_{i}(a_1,a_2)b_i + \epsilon_{i}(a_1,a_2), 
\label{eq:mixedmodel}
\end{equation}
where $\beta$ is an unknown $p$-dimensional parameter, 
$b_i$ is a $q$-dimensional ($q \le p$) 
latent random vector (the random effects) with $\expected{b_i \given L_i}=0$ and
$\epsilon_{i}(a_1,a_2)$ % := (\epsilon_{i1}(a_1,a_2),\ldots, \epsilon_{i,n_i}(a_1,a_2))^\intercal$ 
is the $n_i$-length vector of 
within-subject residual errors with $\expected{\epsilon_i(a_1,a_2)\given L_i}=0$. 
We also assume that $\epsilon_i(a_1,a_2)$ is independent of $b_i$, given $L_i$.
The $n_i \times p$ design matrix $X_i(a_1,a_2)$ %with rows $X_{ij}(a_1,a_2)^\intercal, j=1,\ldots,n_i$ 
depends on the SMART design and a chosen model for the mean, conditional on 
the baseline covariate vector $L_i$. 
The $n_i \times q$ random effects design matrix $Z_{i}(a_1,a_2)$ 
is a function of time, $X_{i}(a_1,a_2)$, and $(a_1,a_2)$ chosen 
so that $Z_{i}(a_1,a_2) b_i $ models subject-specific 
deviations from the mean over time. 
Since $(a_1,a_2)$ indexes the embedded DTRs
and is not a random variable, 
$X_{i}(a_1,a_2)$ and $Z_{i}(a_1,a_2)$ are random 
variables only as a function of $L_i$. 
(In this article we do not treat $t_{ij}$ as a random variable.)
Note that model \eqref{eq:mixedmodel} implies that
$\expected{Y_i(a_1,a_2) \given L_i, b_i} = X_{i}(a_1,a_2)\beta + Z_{i}(a_1,a_2)b_i $
and
$\expected{Y_i(a_1,a_2) \given L_i} = X_{i}(a_1,a_2)\beta$. 

%\subsection{Marginal Mean Modeling}
With model \eqref{eq:mixedmodel}, 
we make primary aim comparisons among embedded DTRs 
based on the linear, parametric marginal model
% $\mu_t(a_1,a_2,L_i;\beta)$ 
for $\expected{Y_{it}(a_1,a_2) \given L_i}$ 
given by 
$%\expected{Y_{it}(a_1,a_2) \given L_i} = 
%\mu_t(a_1,a_2,L_i; \beta) =
 \beta^\intercal X_{it}(a_1,a_2)$,
where $X_{it}(a_1,a_2)^\intercal$ is the row 
of $X_{i}(a_1,a_2)$ evaluated at $t_{ij} = t$. 
%where 
%$\beta$    % =(\beta_1,\ldots,\beta_p)^\intercal$
% is a 
%$p$-dimensional column-vector of unknown parameters. 
%Recall that $L_i$ is a vector of baseline covariates and that
% $(a_1,a_2)$ indexes the embedded DTRs 
%  and is not a random variable.
This is a \emph{marginal} mean model in that $\expected{Y_{it}(a_1,a_2)\given L_i}$
is marginal over both the embedded tailoring variable, $R_i(a_1)$, and 
any other intermediate random variables possibly impacted by $a_1$ or 
$(a_1, a_2)$. 
For the autism SMART, an example marginal mean model 
used previously \citep{lu:2016,almirall2016longitudinal} is 
a piecewise linear model with a knot at week $t_j=12$: 
\begin{align}
%\mu_{t}(a_1,a_2,L_i;\beta) =
\beta^\intercal X_{it}(a_1,a_2) = 
 \beta_0 + t^{ [0,12] } \left(\beta_1  + \beta_2 a_1 \right) + t^{ (12,36] }\left( \beta_3 + \beta_4 a_1 + \beta_5  \indic{a_1=1} a_2 \right) + \beta_6\texttt{age}_i,
 \label{eq:autism-mean-model}
\end{align}
where $\indic{\cdot}$ is the indicator function,
$t^{ [0,12] } = \left( t \indic{t \le 12} + 12\indic{t > 12} \right)$,
$t^{ (12,36] } = (t - 12)\indic{t > 12}$,
and $L_i=\texttt{age}_i$ is the mean-centered age at baseline.
In this case, 
\begin{align*}
X_{it}(a_1,a_2) = [1, t^{ [0,12] }, t^{ [0,12] } a_1, t^{ (12,36] }, 
t^{ (12,36] } a_1, t^{ (12,36] } \indic{a_1=1} a_2, \texttt{age}_i ]^\intercal.
\end{align*}
In this example, the parameters $\beta_2$, $\beta_4$, and $\beta_5 $ have a causal interpretation
and can be used to specify the DTR effect estimands 
of primary interest. An example estimand of primary 
interest may be 
%$\mu_{24}(1,1,L_i; \beta) - \mu_{24}(-1,\cdot, L_i; \beta),$
$
\expected{Y_{i24}(1,1)} - \expected{Y_{i24}(-1, \cdot)} = 12(2(\beta_2+\beta_4)+\beta_5),
$
 an end-of-treatment comparison between the DTR with no AAC, 
  $(a_1, a_2) = (1, 1),$ and the DTR with the highest dose of AAC, 
  $(a_1, a_2) = (-1,\cdot)$.
Other DTR effect estimands are similarly formed via linear combinations of 
$\beta_2$, $\beta_4$, and $\beta_5$. 

%%% MODELING V, V IS CONSEQUENCE OF RANDOM EFFECTS
In addition to specifying $\beta^\intercal X_{it}(a_1,a_2)$ as
a model for $\expected{Y_{it}(a_1,a_2)\given L_i}$, 
 model \eqref{eq:mixedmodel} implicitly defines a 
working model for the marginal covariance 
$V_i(a_1,a_2) := \mathrm{Var}\left(Y_i(a_1,a_2) \given L_i\right)$. 
Since we assume $b_i$ and $\epsilon_i(a_1,a_2)$ are independent given $L_i$,
we have 
$V_i(a_1,a_2) = Z_i(a_1,a_2) \mathrm{Var}(b_i \given L_i )Z_i(a_1,a_2)^\intercal + \mathrm{Var}(\epsilon_i(a_1,a_2) \given L_i)$.
Previously, models for SMARTs with a longitudinal outcome 
involved directly specifying a working model
 for  $V_i(a_1,a_2)$
\citep{seewald2018-repeatedmeasures-power-compareDTRs,
%nahum-shani2018-tutorial-repeatedmeasures-compareDTRs-SMART,
almirall2016longitudinal,
lu:2016}. In contrast, 
the working model for $V_i(a_1,a_2)$ 
 in \eqref{eq:mixedmodel} 
is a consequence of separately modeling within-subject
and between-subject variation via $Z_{i}(a_1,a_2) b_i$ and 
$\epsilon_{i}(a_1,a_2)$. 
Together, the variance and covariance structures specified for 
$b_i$ and $\epsilon_{i}(a_1,a_2)$ imply
 a working model for $V_i(a_1,a_2)$.

\section{Estimation and prediction}
\label{sec:estimation-prototypical}

To derive a set of estimating equations for $\beta$, we 
initially consider the following two distributional assumptions,
which are typical for a 
mixed model like \eqref{eq:mixedmodel}:
\begin{align}
b_i \given L_i &\sim N(0, G)\label{eq:model-assump1-2} \hspace{18pt}
\epsilon_i(a_1,a_2) \given L_i \sim N(0,\sigma^2 I_{n_i}) 
\end{align}
With the addition of the assumptions in \eqref{eq:model-assump1-2}, 
we have $Y_i(a_1,a_2) \given L_i \sim N(X_i(a_1,a_2)\beta, V_i(a_1,a_2))$ 
with $V_i(a_1,a_2)=Z_i(a_1,a_2)G Z_i(a_1,a_2)^\intercal + \sigma^2 I_{n_i}$. 
Based on this distribution for $Y_i(a_1,a_2) \given L_i$, 
 the log-likelihood for a sample % for the potential outcomes
 of $N$ participants under DTR $(a_1,a_2)$ is 
\begin{align}
\begin{split}
&-\frac{1}{2}\sum_{i=1}^N \log\det\left[V_i(a_1,a_2)\right] \\
&\hspace{18pt}
-\frac{1}{2}\sum_{i=1}^N (Y_{i}(a_1,a_2)-X_i(a_1,a_2)\beta)^\intercal V_i(a_1,a_2)^{-1} (Y_{i}(a_1,a_2)-X_i(a_1,a_2)\beta),
\end{split}
\label{eq:po-lik}
\end{align}
In practice, this log-likelihood 
 cannot be maximized since the potential
 outcomes $Y_{i}(a_1,a_2)$ are not observed for all participants 
 under all DTRs in a SMART. 
Instead, we propose a weighted pseudo-likelihood 
based on the observed data collected in a SMART. 
%It will be shown that the estimator for $\beta$ 
%base on the proposed pseudo-likelihood 
%this estimator is consistent and asymptotically Gaussian 
%under a less restrictive set of assumptions 
%given in Theorem~\ref{thm:beta-thm}. 
%We will use these working assumptions to motivate and derive 
%an estimator for $\beta$, but it will be shown that 

\subsection{Pseudo-Likelihood Estimation}
\label{sec:weighted-lik}
The log-likelihood \eqref{eq:po-lik} is a function of the following parameters: 
$\beta$, $\sigma^2$ and the unique parameters in $G$.
We let $\alpha$ denote the vector of unique variance parameters
in $V_i(a_1,a_2) = V_i(a_1,a_2; \alpha)$, including $\sigma^2$. 
For example, if $b_i$ is a scalar random variable and
$Z_{it}(a_1,a_2) = 1$ for all $a_1, a_2$ and $t$, then 
%  $V_i(a_1,a_2)$  has an exchangeable structure and 
  $\alpha = (\sigma^2, \mathrm{Var}(b_i \given L_i))$. 
  For brevity, we often suppress notation indicating 
that $V_i(a_1,a_2) = V_i(a_1,a_2; \alpha)$ depends on $\alpha$.
%Recall that the observed data 
%collected during a SMART consist of the following:
%$L_i$;
%$A_{1i} \in \left\{+1,-1\right\}$, the randomly assigned first-stage treatment;
%$R_i \in \left\{0,1 \right\}$, the binary tailoring variable; 
%%defined as a summary function of first-stage observations;
%$A_{2i} \in \left\{+1,-1\right\}$, the randomly assigned second-stage treatment;
%and $Y_i$, the observed vector of longitudinal outcomes.
Given the observed data in a SMART, defined in section~\ref{sec:notation}, the pseudo-likelihood we use to estimate $\beta$ is 
\begin{align}
%\begin{split}
l(\beta,\alpha)&=
-\frac{1}{2}\sum_{i=1}^N\sum_{a_1,a_2} \tilde{W}_i(a_1,a_2)
 \left( \log\det\left[V_i(a_1,a_2)\right] +
%&\hspace{12pt}
%-\frac{1}{2}\sum_{i}\sum_{a_1,a_2} \tilde{W}_i(a_1,a_2) 
r_i(a_1,a_2)^\intercal V_i(a_1,a_2)^{-1} r_i(a_1,a_2) \right),\label{eq:wrobj}
%\end{split}
\end{align}
where
$r_i(a_1,a_2)  = r_i(a_1,a_2; \beta) = Y_i - X_i(a_1,a_2)\beta$
and
$\tilde{W}_i(a_1,a_2) = I_i^{(a_1,a_2)}(A_{1i},R_i,A_{2i}) W_i^{(a_1,a_2)}(R_i)$.
%W_i^{(a_1,a_2)}(A_{1i}, R_i,A_{2i})\\
%W_i^{(a_1,a_2)}(A_{1i}, R_i,A_{2i})
%W_i^{(a_1,a_2)}(R_i) &=\frac{1}{\prob{A_{1i}=a_1}}\left[R_i + \frac{1}{\prob{ A_{2i} = a_2 \given A_{1i}=a_1, R_i=0}}(1-R_i)\right], 
The indicator $I_i^{(a_1,a_2)}(A_{1i},R_i,A_{2i})$ is equal to one if
and only if the sequence $(A_{1i},R_i,A_{2i})$ is observable under DTR 
 $(a_1,a_2)$. For example, in the autism SMART,
  $I_i^{(a_1,a_2)}(A_{1i},R_i,A_{2i}) = \indic{A_{1i}=a_1}(R_i+(1-R_i)\indic{A_{2i}=a_2})$,
  where $\indic{v}$ equals $1$ if the event $v$ occurs and equals zero otherwise.
  The design-specific weight 
  $W_i^{(a_1,a_2)}(R_i) := \prob{A_{1i} = a_1, A_{2i} = a_2 \given R_i}^{-1}$ 
  is an inverse
  probability weight for the DTR $(a_1,a_2)$ which depends on $R_i$
because second-stage randomization is restricted according to this
binary tailoring variable. In the autism SMART, and in many two-stage designs, 
only individuals with $R_i=0$ are re-randomized, and
$
W_i^{(a_1,a_2)}(R_i) =\frac{1}{\prob{A_{1i}=a_1}}\left[R_i + \frac{1}{\prob{ A_{2i} = a_2 \given A_{1i}=a_1, R_i=0}}(1-R_i)\right].
$
    (When $A_{2i}$ is not defined for a given value of $a_1$, 
  we set $\prob{A_{2i}=a_2 \given A_{1i}=a_1,R_i=0} = 1$.)

Differentiating \eqref{eq:wrobj} with respect to $\beta$ leads to 
the following $p$-dimensional set of estimating equations:
\begin{align}
\sum_{i=1}^N\sum_{a_1,a_2}\tilde{W}_i(a_1,a_2)X_i(a_1,a_2)^\intercal V_i(a_1,a_2; \alpha)^{-1}r_i(a_1,a_2; \beta)=0 ,\label{eq:beta-ee}
\end{align}
with the solution
\begin{align}
\begin{split}
 \hat{\beta}(\alpha) &= \left(\sum_{i=1}^N\sum_{a_1,a_2}\tilde{W}_i(a_1,a_2) X_i(a_1,a_2)^\intercal V_i(a_1,a_2; \alpha)^{-1} X_i(a_1,a_2) \right)^{-1}\\
&\hspace{36pt} \left(\sum_{i=1}^N\sum_{a_1,a_2}\tilde{W}_i(a_1,a_2) X_i(a_1,a_2)^\intercal V_i(a_1,a_2; \alpha)^{-1} Y_i\right).
 \end{split} \label{eq:betahat}
 \end{align}
Substituting $\hat{\beta}(\alpha)$ into \eqref{eq:wrobj}, 
we can obtain estimates of $\beta$ by first computing
$\hat{\alpha}=\arg\max_{\alpha} l(\hat{\beta}(\alpha), \alpha)$
 and then estimating $\beta$ with $\hat{\beta}:=\hat{\beta}(\hat{\alpha})$.
In the following theorem we derive the asymptotic properties of $\hat{\beta}$.

\begin{theorem}
\label{thm:beta-thm}
%Let $\beta^*$ be the true, marginal mean parameter value in 
%the mixed effects model \eqref{eq:mixedmodel}. 
Define 
$
U_i(\beta, \alpha) = \sum_{a_1,a_2}\tilde{W}_i(a_1,a_2)
    		X_i(a_1,a_2)^\intercal V_i(a_1,a_2 ; \alpha)^{-1}
				(Y_i-X_i(a_1, a_2) \beta)
$
and let $\hat{\beta}(\alpha)$ be the solution to $\sum_i U_i(\beta, \alpha)=0$ 
given in \eqref{eq:betahat}. 
%Let $\hat{\alpha}
Assume the following:
\renewcommand{\theenumi}{\roman{enumi}}%
\begin{enumerate}
\item Correctly specified marginal model: $\expected{Y_{i}(a_1,a_2) \given L_i} = X_i(a_1,a_2) \beta^*$
\item Sequential randomization: $Y_i(a_1,a_2)$ is independent of $A_{1i}$ given $L_i$;
$R_i(a_1)$ is independent of $A_{1i}$ given $L_i$;
and 
$Y_i(a_1,a_2)$ is independent of $A_{2i}$ given $(A_{1i}, R_i, L_i)$.
\item Consistency: 
$R_i = R_i(A_{1i}) = \sum_{a_1} \indic{A_{1i}=a_1} R_i(a_1)$
and
\begin{align*}
Y_i &= R_iY_i(A_{1i}) + (1-R_i)Y_i(A_{1i},A_{2i})\\
 &= \sum_{a_1} \indic{A_{1i}=a_1}R_i(a_1)Y_i(a_{1}) + \sum_{a_1, a_2} \indic{A_{1i}=a_1}\indic{A_{2i}=a_2}(1-R_i(a_1))Y_i(a_{1}, a_{2}),
 \end{align*}
where $ R_iY_i(A_{1i}) : = R_i Y_i(A_{1i}, a_2)$ for all $a_2$.
\item Positivity: $\prob{A_{1i} = a_1} >0$ and $\prob{A_{2i}=a_2 \given A_{1i}, R_i=0}>0$ for any 
$a_1$, $a_2$. 
\item Regularity conditions: For any given $\beta$, $\hat{\alpha} = \hat{\alpha}(\beta)$ converges to some $\alpha^*$ at $\sqrt N$ rate, and
\begin{align*}
& \sup_\beta\left\|\frac{1}{N}\sum_iU_i(\beta, \hat\alpha (\beta)) - \lim_{N\to\infty}\frac{1}{N}\sum_i \expected{U_i(\beta, \alpha^* (\beta))\given L_i}\right\| \stackrel{P}{\to} 0.
\end{align*}
\end{enumerate}
Then 
$\hat{\beta}(\hat{\alpha})$ is consistent for $\beta^*$ and $\sqrt{N}(\hat{\beta} - \beta^*)$ 
has an asymptotic $N(0, J^{-1} I J^{-1})$ distribution, 
where 
$I = \lim_{N\to\infty}\frac{1}{N}\sum_i\expected{U_i(\beta^*,\alpha^*)U_i(\beta^*,\alpha^*)^\intercal \given L_i}$ and 
\begin{align*}
%I &= \lim_{N\to\infty}\frac{1}{N}\sum_i\expected{U_i(\beta^*,\alpha^*)U_i(\beta^*,\alpha^*)^\intercal \given L_i},\\%\,\, U_i(\beta, \alpha) = \sum_{a_1,a_2} \tilde{W}_i(a_1,a_2) X_i(a_1,a_2)^\intercal V_i(a_1,a_2; \alpha)^{-1}r_i(a_1,a_2),\\
J &= \lim_{N\to\infty}\frac{1}{N}\sum_i\expected{\sum_{a_1,a_2} \tilde{W}_i(a_1,a_2) X_i(a_1,a_2)^\intercal V_i(a_1,a_2; \alpha^*)^{-1} X_i(a_1,a_2) \given L_i}.
\end{align*}
\end{theorem}
The diagonal entries of $\frac{1}{N}\hat{J}^{-1}\hat{I}\hat{J}^{-1}$ provide approximate standard errors for $\hat{\beta}$, where 
${\hat{I} = \frac{1}{N} \sum_{i} \hat{U}_i\hat{U}_i^\intercal}$, 
$\hat{U}_i:= U_i(\hat{\beta},\hat{\alpha}) = \sum_{(a_1,a_2)}\tilde{W}_i(a_1,a_2)
    		X_i(a_1,a_2)^\intercal \hat{V}_i(a_1,a_2 ; \hat{\alpha})^{-1}
				(Y_i-X_i(a_1,a_2) \hat{\beta})$, and 
\begin{align*}
%\hat{U}_i:= U_i(\hat{\beta},\hat{\alpha}) &= \sum_{(a_1,a_2)}\tilde{W}_i(a_1,a_2)
   % 		X_i(a_1,a_2)^\intercal \hat{V}_i(a_1,a_2 ; \hat{\alpha})^{-1}
	%			(Y_i-X_i(a_1,a_2) \hat{\beta}),\\
    \hat{J} &= \frac{1}{N} 
    		\sum_i \sum_{a_1,a_2} \tilde{W}_i(a_1,a_2) 
				X_i(a_1,a_2)^\intercal \hat{V}_i(a_1,a_2 ; \hat{\alpha})^{-1} 
					X_i(a_1,a_2).
\end{align*}

The proof of Theorem~\ref{thm:beta-thm} is given in the appendix.
Note that assumption (ii) and (iv), above, will be satisfied by design of the SMART, 
while 
%(i) states that the mean model is correctly specified and (v) is a regularity condition 
%for the estimating function $\frac{1}{N}\sum_i U_i(\beta,\hat{\alpha}(\beta))$. 
assumption (iii) connects the observed data to the potential outcomes. % $Y_i(a_1,a_2)$ and $R_i(a_1)$. 
Theorem~\ref{thm:beta-thm}
does not require the two assumptions in \eqref{eq:model-assump1-2} to be true. 
%although assumption (i), for example, is a consequence of \eqref{eq:margdist-y}. 
These standard distributional assumptions were used only to motivate 
the pseudo-likelihood and set of estimating equations which led to an estimator 
for $\beta$.
%In particular, the random effects can be misspecified without affecting 
%the conclusion of Theorem~\ref{thm:beta-thm}. 
%This is analogous to the well-known property of 
%GEE in which the working correlation matrix may be misspecified without 
%inducing bias in estimating $\beta$.

Given $V_i(a_1,a_2)$, the estimating equation 
\eqref{eq:beta-ee} is identical, with slight changes in notation,
 to the estimating equation 
 % (3)
 in \citet{lu:2016} for the parameters of the marginal mean model. 
% In our case this estimating equation was motivated by 
 %maximum likelihood estimation of \eqref{eq:mixedmodel} 
 %with the added assumptions of \eqref{eq:model-assump1-2}. 
Estimation of $\beta$ in \citet{lu:2016} differs from our approach primarily
in its modeling and estimation procedure for 
$V_i(a_1,a_2) = \mathrm{Var}(Y_i(a_1,a_2) \given L_i)$. 
In \citet{lu:2016}, the form of $V_i(a_1,a_2)$ (e.g. autoregressive) is 
proposed by the data analyst and an estimate of $V_i(a_1,a_2)$ is
obtained via the method of moments.
 In our case the form of $V_i(a_1,a_2)$ is a result of  
 specifying $Z_{i}(a_1,a_2)b_i$ and the 
 variance-covariance of $\epsilon_{i}(a_1,a_2)$ and $b_i$, 
 while the estimate of $V_i(a_1,a_2)$ is computed 
 by maximizing a weighted pseudo-likelihood.
 
 As in \citet{lu:2016}, Theorem~\ref{thm:beta-thm} implies that 
 $\hat{\beta}$ is consistent for $\beta$ and has an asymptotic Gaussian distribution, 
 regardless of whether $\hat{\alpha}$ converges to the true value of $\alpha$ in model~\eqref{eq:mixedmodel}. 
 This means that the random effects structure can be misspecified 
 and the estimator $\hat{\beta}$ will remain unbiased. 
 However, the simulation results in Section~\ref{sec:sims}
 show that specifying a random effects structure which more closely
 models the true subject-to-subject variation in $Y_i(a_1,a_2)$
 can lead to greater efficiency in estimating $\beta$. 
 Before demonstrating the performance of our estimator
 in simulation studies, we propose a method for
predicting the value of $b_i$ in model~\eqref{eq:mixedmodel} and hence
predicting subject-specific trajectories for the
primary outcome in a SMART.

\subsection{Random Effects Prediction}
\label{sec:blup}
The estimator for $\beta$ derived above is all that is necessary for
primary aim comparisons among the DTRs embedded in a SMART. 
Recall that a secondary motivation for using linear mixed models 
%to analyze SMARTs 
is the prediction of subject-specific outcome 
trajectories under specific DTRs. In this section we propose
a method of predicting $b_i$ in \eqref{eq:mixedmodel} 
using the weighted pseudo-likelihood in \eqref{eq:wrobj}. 

In Theorem~\ref{thm:beta-thm} 
we do not require knowledge of 
$V_i(a_1,a_2)$ or the distributions of $\epsilon_i(a_1,a_2)$ and 
$b_i$. 
% is assumed to prove that $\hat{\beta}(\hat{\alpha})$
% is consistent and has an asymptotic Gaussian distribution. 
 To predict $b_i$, however, we assume that the
 distributional assumptions in \eqref{eq:model-assump1-2} 
are true in the population of potential outcomes.
 Specifically, under model \eqref{eq:mixedmodel},
 assuming \eqref{eq:model-assump1-2}, 
%$b_i \given L_i \sim N(0,G)$,  $\epsilon_i(a_1,a_2) \given L_i \sim N(0,\sigma^2 I_{n_i})$,
% and independence between $b_i$ and $\epsilon_i(a_1,a_2)$, given $L_i$, 
 ${(Y_{i}(a_1,a_2)^\intercal, \, b_i^\intercal ) \given L_i}$ has 
a multivariate Gaussian distribution, which implies that %, and standard calculations 
%    \begin{align*}
%    \mathrm{Var}\left(\left[\begin{array}{c}
%    Y_{i}(a_1,a_2)\\b_i
%    \end{array}\right] \given L_i \right)
%     &=
%     	\left[ \begin{array}{cc}
%	    V_i(a_1,a_2) & Z_i(a_1,a_2)G\\
%	    G Z_i(a_1,a_2)^\intercal & G
%    \end{array}\right].
%    \end{align*}
%In this case, standard properties of the multivariate Gaussian distribution imply that 
\begin{align}
	    b_i \given Y_{i}(a_1,a_2), L_i &\sim 
    		N\left(G Z_i(a_1,a_2)^\intercal  
		V_i(a_1,a_2)^{-1}(Y_{i}(a_1,a_2)-X_i(a_1,a_2)\beta),  
			\, \Sigma_{b\given Y(a_1,a_2)}\right), \label{eq:bposterior}
%\expected{b_i \given Y_{i}(a_1,a_2), L_i } &= G Z_i(a_1,a_2)^\intercal  
%		V_i(a_1,a_2)^{-1}(Y_{i}(a_1,a_2)-X_i(a_1,a_2)\beta). \label{eq:bposterior}
\end{align}
where 
$\Sigma_{b \given Y(a_1,a_2)} = \mathrm{Var}\left(b_i \given Y_{i}(a_1,a_2), L_i \right)$. 
If all potential outcomes $Y_{i}(a_1,a_2)$ were observed for
 each participant, a plug-in estimator of
  $\expected{b_i \given Y_{i}(a_1,a_2), L_i}$ based on 
  \eqref{eq:bposterior} would serve as a prediction of 
  %the $i$th participant's random effects 
  $b_i$. 
  Instead, motivated by \eqref{eq:bposterior}, we propose the following:
    \begin{align}
    \begin{split}
    \hat{b}_i(\alpha, \beta) &= 
    		\arg\max_{b_i} -\frac{1}{2}\sum_{a_1,a_2}
			\tilde{W}_i(a_1,a_2)
				\left(b_i-G Z_i(a_1,a_2)^\intercal 
				V_i(a_1,a_2)^{-1}(Y_i-X_i(a_1,a_2)\beta)\right)^\intercal \\
    &\hspace{84pt} \Sigma_{b \given Y}^{-1}\left(b_i-G Z_i(a_1,a_2)^\intercal
    	V_i(a_1,a_2)^{-1}(Y_i-X_i(a_1,a_2)\beta)\right)
    \end{split}\\[6pt]
&= 
    	\frac{\sum_{a_1,a_2}\tilde{W}_i(a_1,a_2)
		GZ_i(a_1,a_2)^\intercal V_i(a_1,a_2; \alpha)^{-1}
		(Y_i-X_i(a_1,a_2)\beta)}{\sum_{a_1,a_2}
		\tilde{W}_i(a_1,a_2)}. \label{eq:blup-i}
    \end{align}
In practice, the predictions for each participant 
are obtained by substituting the estimates $\hat{\alpha}$ 
and $\hat{\beta}$ into \eqref{eq:blup-i}, so that 
$\hat{b}_i := \hat{b}_i(\hat{\alpha}, \hat{\beta})$. 
(Recall that estimates for the entries of $G$ are
given by some of the components of $\hat{\alpha}$.)
%    \begin{align}
%    \hat{b}_i: = 
%    	\hat{b}_i(\hat{\alpha}, \hat{\beta}) &= 
%	\frac{\sum_{a_1,a_2}\tilde{W}_i(a_1,a_2)
%	\hat{G}Z_i(a_1,a_2)^\intercal \hat{V}_i(a_1,a_2; \hat{\alpha})^{-1}
%	(Y_i-X_i(a_1,a_2)\hat{\beta})}{\sum_{a_1,a_2}\tilde{W}_i(a_1,a_2)} \label{eq:blup-hat}.
%    \end{align}
This predictor can be regarded as an empirical Bayes predictor for $b_i$ 
\citep{skrondal2009prediction, carlin2000empirical} 
with weights that adjust for the probability of observing
responders and slow responders under each 
embedded DTR.

\section{Simulations studies}
\label{sec:sims}
Next we use simulation studies to evaluate the 
ability of our mixed effects model to estimate 
causal estimands of primary interest when comparing
embedded DTRs in a SMART.
We also compare our mixed model estimator to 
the GEE-like estimators discussed in 
\citet{lu:2016} and
  \citet{seewald2018-repeatedmeasures-power-compareDTRs}. 
%  in which the method of moments is used to estimate a working
%   variance-covariance for the repeated measures, which is then
%    used as an estimator for $V_i(a_1,a_2)$ when solving equation \eqref{eq:beta-ee}. 

Data were generated from a hypothetical SMART with
two treatment stages, two treatment options 
for all participants in stage one, and two treatment options
for slow responders in stage two, leading to four
embedded DTRs with $a_1,a_2 \in \left\{1,-1\right\}$.
This is a common SMART design 
\citep[e.g.]{naar2015sequential, august2014being} 
and is different 
from the autism SMART in 
Figure~\ref{fig:exampleSMART_autism}, 
%which has three embedded DTRs: 
in which slow responders from only one of 
the stage-one treatment arms were randomized 
at the start of the second stage. %, leading to three embedded DTRs. 
In a given simulation replicate, potential outcomes 
were generated according to
\eqref{eq:sim-gen}, below, and 
observed data were obtained from these potential 
outcomes via 
% sequential 
randomizations satisfying 
 assumptions (ii) and (iii) in Theorem~\ref{thm:beta-thm}.
 All simulated participants were randomized with equal probability 
 to either $A_{1i} = 1$ or $A_{1i}=-1$, and only slow responders
 were randomized to $A_{2i} = 1$ or $A_{2i}=-1$ with equal probability. 
%Random intercepts and slopes induced within-person correlation
%in the longitudinal potential outcomes, and the within-person residual
%errors were independent across time. 

The potential outcomes in these simulation studies were generated
 from the following piecewise linear model:
 \begin{align}
    \begin{split}
    &Y_{it}(a_1,a_2) =    \theta_0 + \indic{t \le \kappa} t(\theta_1+\theta_2 a_1) + \indic{t > \kappa} \kappa (\theta_1+\theta_2 a_1)\\
   	&\hspace{6pt}+ \indic{t > \kappa}(t-\kappa)(\theta_3+\theta_4 a_1 + (\theta_5 a_2+\theta_6 a_1a_2)(1-R_i(a_1)))\\
    	&\hspace{6pt}+ \indic{t > \kappa}(t-\kappa)(\psi^{(1)}\indic{a_1=1} + \psi^{(-1)}\indic{a_1=-1})\left[R_i(a_1)-\prob{R_i(a_1)=1 \given L_i}\right]\\
            &\hspace{6pt} +\theta_7 L_i + \gamma_{0i} + \gamma_{1i} t + \epsilon_{it},
    \end{split}\label{eq:sim-gen}
    \end{align}
where $R_i(a_1) = \indic{Y_{i \kappa}(a_1) -\theta_7 L_i > c}$; $c = 1.1$;
 %$\theta_7 = -0.2$; $
$(\gamma_{0i}, \gamma_{1i})^\intercal \given L_i \sim N(0, \Gamma);$ 
and $\epsilon_{it} \given L_i \sim N(0, \tau^2)$ with $\tau^2 = 1$.
%$t \in \left\{0, 0.5, 1.5, 2, 2.25, 2.5, 3\right\}$; and $\kappa = 2$. 
The binary tailoring variable $R_i(a_1)$
is a function of the potential outcome
at the end of the first treatment stage, 
and the fixed value of $c$ means that $\prob{R_i(a_1)=1 \given L_i}$ 
 varies as a function of $a_1$. 
 The parameters $\psi^{(1)}$ and $\psi^{(-1)}$ induce a marginal association
 between $R_i(a_1)$ and second-stage outcomes. 
The random intercepts and slopes, $\gamma_{0i}$ and $\gamma_{1i}$, induce 
within-person correlation,  %in the longitudinal potential outcomes, 
and the residual errors $\epsilon_{it}$ were generated independently across $i$ and $t$. 
The scalar random variable $L_i$ is a binary baseline covariate, and
the knot $\kappa$ represents the time when the first treatment stage ends. 
In all simulations, half of the participants were assigned
 $L_i=1$ and half were assigned $L_i=-1$. 
Under \eqref{eq:sim-gen}, the marginal mean can be expressed as follows: 
%linear function of time: 
    \begin{align}
    \begin{split}
&\expected{Y_{it}(a_1,a_2) \given L_i}= 
     \beta^\intercal X_{it}(a_1,a_2)  = \beta_0 +\indic{t \le \kappa}t (\beta_1+\beta_2 a_1) + 
    \indic{t > \kappa}\left(\kappa(\beta_1+\beta_2 a_1)\right)\\
    	& \hspace{24pt}+
	\indic{t > \kappa} \left((t-\kappa)(\beta_3+\beta_4 a_1+\beta_5 a_2 + \beta_6 a_1a_2)
	\right) + \beta_7 L_i + \gamma_{0i} + \gamma_{1i} t + \epsilon_{it},
    \end{split}  \label{eq:sim-mmean}
    \end{align}
and causal estimands of primary interest are expressed as
functions of $\beta_2, \beta_4, \beta_5$, and $\beta_6$. 
Further details of this generative model 
are given in the appendix.

Below we present the results of two simulation studies 
which differ in whether or not they misspecify the marginal variance and distribution
of $Y_i(a_1,a_2)$. In both cases, the 
linear model for $\expected{Y_i(a_1,a_2) \given L_i}$ is 
correctly specified. 
We report estimation performance for the end-of-study contrast 
$\expected{Y_{i 3}(1,-1) \given L_i} - \expected{Y_{i 3}(-1,-1)\given L_i} = 
2\kappa\beta_2 + 2(3-\kappa)\beta_4 - 2(3-\kappa)\beta_6,$
and we chose simulation parameters so that this contrast had the largest magnitude 
among any pairwise contrast between embedded DTRs. 
Parameter values for the marginal mean were chosen to achieve 
desired values of the standardized effect size
$d = \frac{\expected{Y_{i 3}(1,-1) \given L_i} - \expected{Y_{i 3}(-1,-1)\given L_i} }{\sqrt{\frac{1}{2}\mathrm{Var}\left(Y_{i3}(1,-1)\given L_i \right) + \frac{1}{2} \mathrm{Var}\left(Y_{i3}(-1,-1) \given L_i\right)}}$.

\subsection{Simulation 1}
The first simulation study verifies that our estimator $\hat{\beta}$ 
is unbiased in large samples and that large-sample confidence interval coverage 
is attained with the standard errors based
on Theorem~\ref{thm:beta-thm}.
This is accomplished in the ideal setting in 
which the probability distribution of $Y_i(a_1,a_2) \given L_i$ 
can be correctly specified using our proposed mixed model.
%in equation~\eqref{eq:mixedmodel}. 
%
%parameter values for the generative model \eqref{eq:sim-gen} so that the
In general, 
$Y_i(a_1,a_2) \given L_i$ in \eqref{eq:sim-gen} 
follows a Gaussian mixture distribution with mixing probability
$\prob{R_i(a_1)=0 \given L_i}$.
%under generative model \eqref{eq:sim-gen}.
However, in this simulation study we choose
  $0 = \theta_5 = \theta_6 = \psi^{(1)} = \psi^{(-1)}$, 
  so that 
  $Y_{it}(a_1,a_2)-\expected{Y_{it}(a_1,a_2) \given L_i} = \gamma_{0i} + \gamma_{1i} t +\epsilon_{it}$
and the distribution of $Y_i(a_1,a_2)\given L_i$ is
the same as the marginal distribution 
specified in the following mixed model:
    \begin{align}
%    \begin{split}
    Y_{it}(a_1,a_2) \given b_{0i}, b_{1i}, L_i &\sim N(\beta^\intercal X_{it}(a_1,a_2) + b_{0i}+b_{1i}t, \sigma^2),&
    (b_{0i}, b_{1i})^\intercal \given L_i &\sim N(0, G),
    %\end{split}
    \label{eq:sim-mmest-slopes}
    \end{align} 
where $\beta^\intercal X_{it}(a_1,a_2)$ is the linear parametrization of the mean in equation \eqref{eq:sim-mmean}.
We compared this ``slopes and intercepts'' mixed model, in which the joint distribution of $Y_i(a_1,a_2)\given L_i$
is correctly specified,
to
an ``intercepts only'' mixed model,
    \begin{align}
%    \begin{split}
    Y_{it}(a_1,a_2) \given b_{0i}, L_i &\sim N(\beta^\intercal X_{it}(a_1,a_2)+ b_{0i}, \sigma^2), &
    b_{0i} \given L_i &\sim N(0,\mathrm{Var}\left(b_{0i}\right)),
%    \end{split}
    \label{eq:sim-mmest-intercepts}
    \end{align}
in which $\mathrm{Var}(Y_i(a_1,a_2)\given L_i)$ is misspecified. 
We use different notation for the random effects and variance parameters in 
\eqref{eq:sim-mmest-slopes}--\eqref{eq:sim-mmest-intercepts} than we do in \eqref{eq:sim-gen} 
to distinguish models used for estimation from 
the true, data-generating probability distribution. 
In this simulation study %, for the random effects in \eqref{eq:sim-gen}, we set 
we set 
 $\mathrm{Var}\left(\gamma_0\right)= 0.8, 
 \mathrm{Var}\left(\gamma_1\right)= 1$ and 
 $\mathrm{Cov}(\gamma_0, \gamma_1) = -0.2$. 

%in which the random effects structure and variance parameters
%may or may not be equivalent to those of 
%\eqref{eq:sim-mmest-slopes}--\eqref{eq:sim-mmest-intercepts}.

 Table~\ref{tab:sim1} contains the bias and 
 standard deviation of the point estimates, 
 the mean of the approximate standard errors, 
 the coverage probability for a 95-percent confidence interval,
 and the root mean squared error (RMSE) computed from 1,000 simulation replicates. 
In large samples, the bias is approximately two orders of magnitude smaller than the 
standard deviation of the point estimates, 
confirming that the mixed model 
estimator is unbiased for the linear mean parameters in \eqref{eq:sim-mmean}. 
The standard errors based on Theorem~\ref{thm:beta-thm} provide 
confidence interval coverage close to the nominal level in large samples.
In addition, note that the intercepts only mixed model, 
which misspecifies 
$\mathrm{Var}(Y_i(a_1,a_2) \given L_i)$, does not introduce bias in large samples. 
Instead, the estimator is slightly less efficient than the slopes and
 intercepts model, in which both the mean and covariance of $Y_i(a_1,a_2)$ are correctly specified. 

\subsection{Simulation 2}
\label{sec:sim2}
In this second simulation, we 
investigate whether the estimator
$\hat{\beta}$ is unbiased in large samples, and 
whether this estimator can provide 
efficiency gains relative to existing estimators, 
in a more realistic scenario
in which it is not possible to correctly specify the distribution
of $Y_i(a_1,a_2) \given L_i$ using
model \eqref{eq:mixedmodel}. 
Data were again generated from model \eqref{eq:sim-gen}, but
 the coefficients $\theta_5, \theta_6, \psi^{(1)}$ and $\psi^{(-1)}$ 
 were nonzero and therefore $Y_i(a_1,a_2) \given L_i$ 
 had a marginal Gaussian mixture distribution. 
%All of the estimation methods compared here correctly 
%specify the linear mean model \eqref{eq:sim-gen}. 

In addition to the mixed models 
\eqref{eq:sim-mmest-slopes}--\eqref{eq:sim-mmest-intercepts}, 
we also report estimation performance of the
GEE-like estimator of \citet{lu:2016} and 
\citet{seewald2018-repeatedmeasures-power-compareDTRs}. 
%in which
%only the marginal mean is assumed to be correctly specified and no further 
%distributional assumptions are made about $Y_i(a_1,a_2) \given L_i$.
With these GEE estimators, a working model for $V_i(a_1,a_2)$ (e.g. exchangeable) 
is specified directly and the method of moments is used to estimate the parameters in this working model. 
A complete description of our implementation of these GEE estimators 
%including the method of moments estimators for $V_i(a_1,a_2)$ 
is given in the appendix. 
In this simulation study, 
all of the models used for estimation % of the end-of-study contrast
 correctly specify the linear model $\beta^\intercal X_{it}(a_1,a_2)$, but none
of them correctly specify the marginal covariance or the distribution of $Y_i(a_1,a_2) \given L_i$.
 
%The GEE-like estimators of \citet{lu:2016} and 
%\citet{seewald2018-repeatedmeasures-power-compareDTRs}
%use working models 
%These estimators differ primarily in the implied model for 
%$V_i(a_1,a_2) = \mathrm{Var}\left(Y_i(a_1,a_2) \given L_i\right)$ 
% (e.g. an exchangeable structure) and in how
%  they compute an estimate of $V_i(a_1,a_2)$. 
%  The LMM models use the weighted pseudo-likelihood 
%  described in Section~\ref{sec:weighted-lik} while the GEE-like estimators use the method of moments.

%
%The GEE-like estimator of \citet{seewald2018-repeatedmeasures-power-compareDTRs} and \citet{lu:2016} was implemented as follows. 
%First we compute
%\begin{align*}
%\hat{\beta}^{(0)} =  \left(\sum_i\sum_{a_1,a_2}\tilde{W}_i(a_1,a_2)X_i(a_1,a_2)^\intercal X_i(a_1,a_2)\right)^{-1}\left(\sum_i\sum_{a_1,a_2}\tilde{W}_i(a_1,a_2)X_i(a_1,a_2)^\intercal Y_i\right)
%\end{align*}
%and the $n_i$-length vector of residuals 
%$r_i^{(0)}(a_1,a_2) = Y_i-X_i(a_1,a_2)\hat{\beta}^{(0)}$ for all $i$ and 
%$(a_1,a_2)$. These residuals are then used to compute 
%a method of moments estimator for $V_i(a_1,a_2)$ 
%according to a proposed working model, such as exchangeable 
%or autoregressive. 
%Finally, we use the method of moments estimate for 
%$V_i(a_1,a_2)$ in equation \eqref{eq:betahat} to obtain $\hat{\beta}$. 
%Further details on the method of moments estimators for $V_i(a_1,a_2)$ are given in the supplementary materials (available online).

Table~\ref{tab:sim2} compares the two mixed models and 
the GEE estimator with 
exchangeable, unstructured, and independence working models for $V_i(a_1,a_2)$
 in their ability to estimate the end-of-study contrast with standardized effect size $d \approx 0.5$. 
The magnitude of the bias relative to the standard deviation in Table~\ref{tab:sim2} 
indicates that all of these estimators are unbiased in large samples.
%However, 
%In addition to the bias, standard deviation, average standard error estimate,
%and coverage probability, 
%We report RMSE in Table \ref{tab:sim2} 
%as a fraction of the smallest RMSE for a fixed sample size. 
%The slopes and intercepts mixed model had the lowest RMSE 
%for each fixed sample size, and the GEE independence estimator 
%has an RMSE about 30 percent larger than that of 
%the slopes and intercepts model at all three report sample sizes. 
%The standard error estimates at smaller sample sizes seem to underestimate
%the true sampling variability in t
%These results also
%suggest that the estimated standard errors may be too small relative
%to the true sampling variability, especially with smaller sample sizes. 

While none of the estimation models in this simulation study 
correctly specify $V_i(a_1,a_2)$, 
we can see in Table~\ref{tab:sim2} that efficiency 
(measured by RMSE) 
%in estimating the end-of-study contrast 
is improved 
by a working model %for $V_i(a_1,a_2)$ 
which more closely
resembles the true marginal covariance. 
Here we report RMSE % in Table \ref{tab:sim2} 
as a fraction of the smallest RMSE for a fixed sample size. 
To measure the performance of each working model for $V_i(a_1,a_2)$,
we report 
$\frac{\left\|V_{\text{true}} - \expected{\hat{V}}\right\|}{\left\|V_{\text{true}}\right\|}$,
the relative error in the Frobenius norm between 
 $V_{\text{true}} := \frac{1}{4} \sum_{a_1,a_2} V_i(a_1,a_2)$, 
 the true average covariance matrix of
  $Y_i(a_1,a_2)$ according to the generative model, and
the simulation-based estimate of 
$\expected{\hat{V}} := \frac{1}{4}\sum_{a_1,a_2}\expected{ \hat{V}_i(a_1,a_2; \hat{\alpha})}$,
the large-sample average covariance matrix implied by the estimation model.
The slopes and intercepts mixed model had both 
the lowest relative error in estimating $V_{\text{true}}$ 
and the lowest RMSE for each fixed sample size.
For these estimators, RMSE decreases as the 
working model for $V_i(a_1,a_2)$ improves.
%had the lowest RMSE 
%for each fixed sample size, and the GEE independence estimator 
%has an RMSE about 30 percent larger than that of 
%the slopes and intercepts model,
%% at all three report sample sizes. 
%The relative error in estimating $V_{\text{true}}$ 
This simulation study suggests that 
if the separate specification of between-person 
and within-person variation in a mixed effects model 
leads to improved modeling of the marginal covariance, 
we can expect efficiency gains over GEE-like approaches 
when comparing embedded DTRs.

\section{Application}\label{sec:application}
\begin{figure}[!p]
\centering
\includegraphics[width=\textwidth]{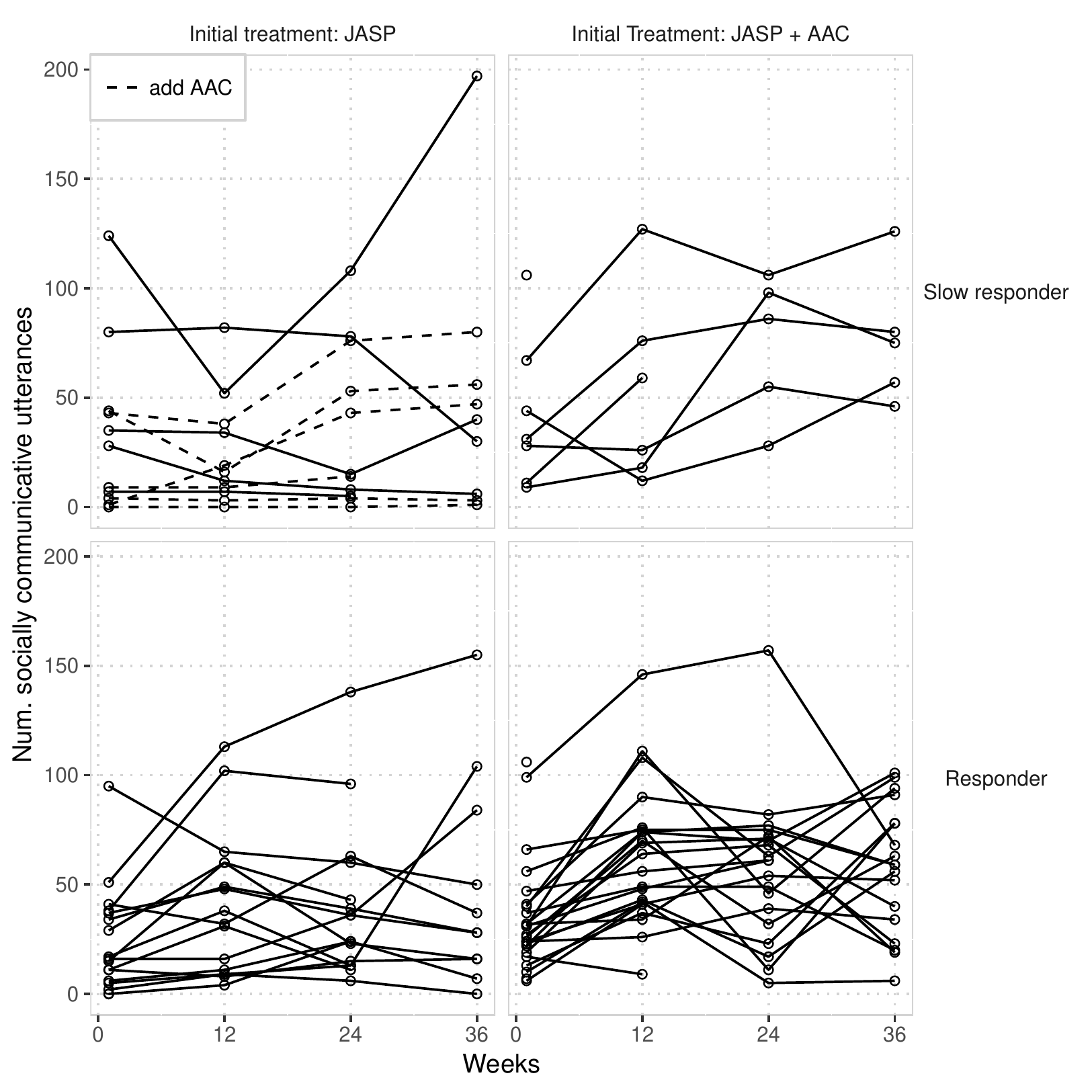}
\caption{Measured number of socially communicative utterances for the $N=61$ children in the autism SMART. Responders to either first-stage treatment continued that treatment. Dashed lines in the upper-left panel correspond to slow responders to initial JASP who were randomly assigned to receive JASP+AAC in the second stage. All other slow responders received an intensified version of the initial treatment.}
\label{fig:autism-raw}
\end{figure}

Finally, we demonstrate our mixed model using the autism SMART
of \citet{kasari2014communication}. 
%This study sought to understand
%how a speech-generating device (AAC) can be incorporated into
%a specific behavioral treatment (JASP).
Our goal here is to compare the three embedded
DTRs based on changes in communication outcomes for the children receiving 
each DTR.
 Figure~\ref{fig:autism-raw} displays the measured primary outcome, 
 the number of socially communicative utterances, 
 for each participant in this study 
 at baseline and at weeks 12, 24, and 36. 
For the marginal mean, we specified the piecewise linear model \eqref{eq:autism-mean-model}, 
and we specified random intercepts as the random effects structure. 

\begin{figure}[!p]
\centering
\includegraphics[width=0.66\textwidth]{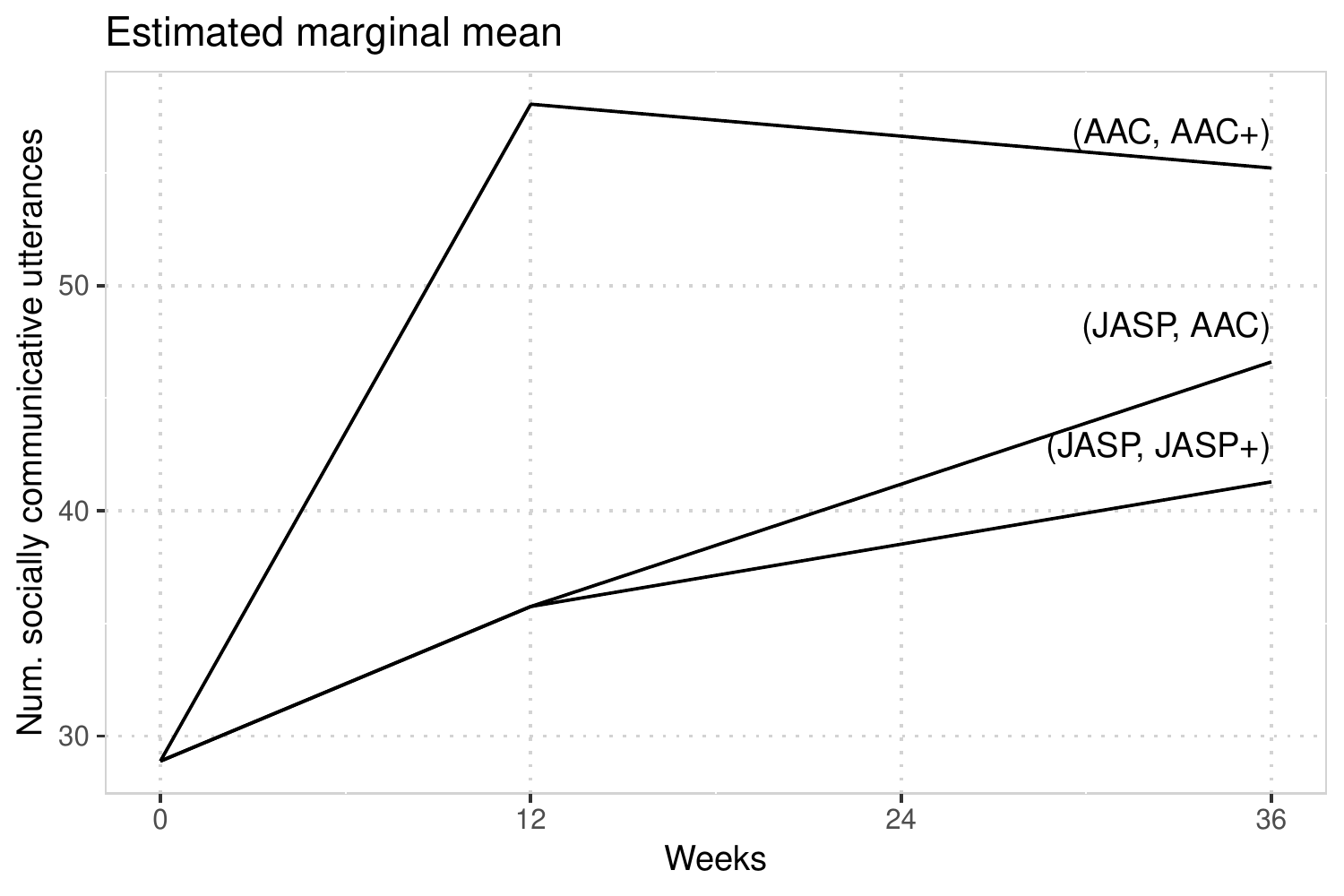}
\caption{Estimated marginal mean, at age $6.3$ (the average age of participants in the study), for the three treatment regimens in the autism SMART.}
\label{fig:muhat}
\end{figure}

\begin{figure}[!p]
\centering
\includegraphics[width=\textwidth]{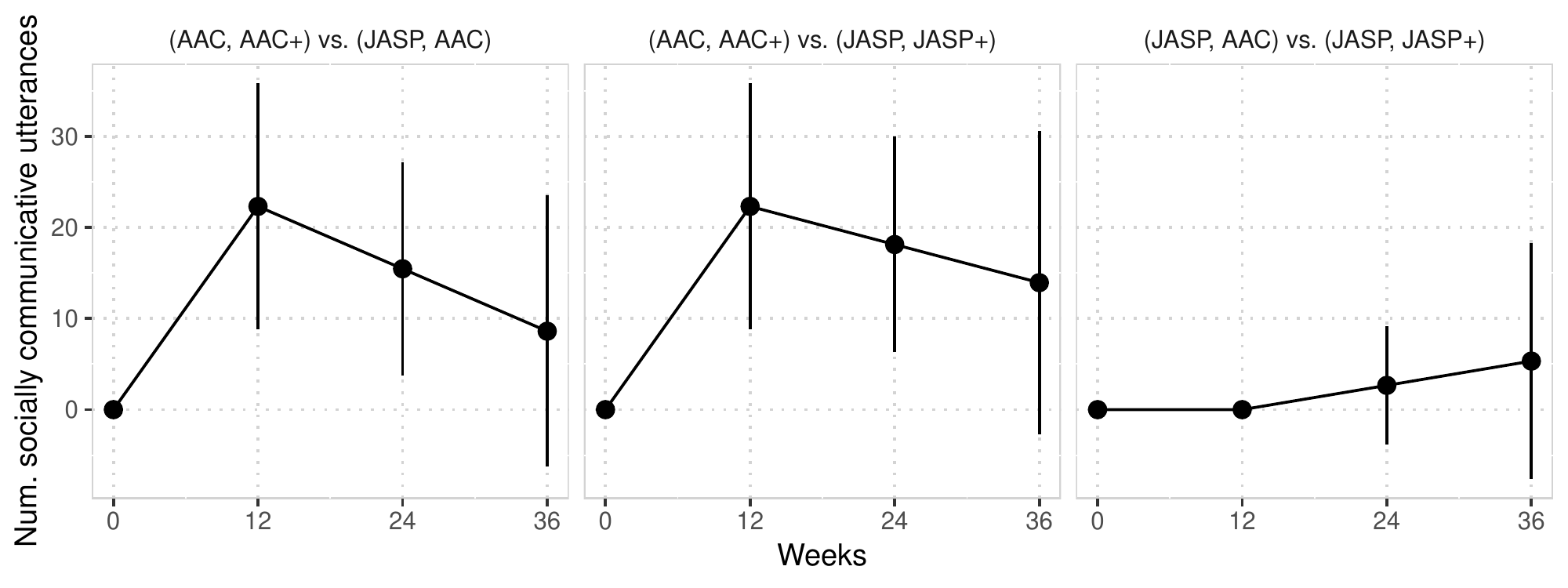}
\caption{Pairwise contrasts with 95-percent pointwise confidence intervals for the autism SMART.}
\label{fig:pairwise-contrasts}
\end{figure}

The parameter vector $\beta = (\beta_0,\ldots,\beta_6)$ was estimated 
as described in Section~\ref{sec:weighted-lik} using 
widely available software for linear mixed models 
\citep{lme4-bates} applied to a restructured version of the observed data. 
This software implementation is described in greater detail
 in the appendix.
The estimates of $\alpha$ and 
$\beta$ obtained in this manner were then used to compute
estimated standard errors as described in Section~\ref{sec:weighted-lik}. 
Table~\ref{tab:betahat} displays
 the estimated coefficients in this model with
  95-percent confidence intervals, and 
 Figure~\ref{fig:muhat} 
displays the estimated marginal mean for each DTR at each time point, 

To understand whether we have evidence that
communication outcomes differ among
children receiving each of these
DTRs, we performed an 
``omnibus'' test of whether the three DTRs differ at all. 
We tested the hypothesis that 
the area under the curve (AUC) for the marginal 
mean is the same across all three DTRs, 
which, in this case, is equivalent to testing 
%Specifically, we tested the hypothesis that 
%the area under the curve (AUC) for the marginal mean
%under each DTR is equal:
%$
%H_0: \text{AUC}^{(1,1)} = \text{AUC}^{(1,-1)} = \text{AUC}^{(-1,\cdot)}.
%$
%With the piecewise linear model given above, 
%the AUC for each DTR is %the sum of the areas of two trapezoids and
% equal to a linear combination of the parameter vector $\beta$.
%In other words, the omnibus hypothesis is 
 $H_0: M \beta = 0$ for a constant matrix $M$.
%> Lmat_omnibus
%     [,1] [,2] [,3] [,4] [,5] [,6] [,7]
%[1,]    0    0    0    0    0  576    0
%[2,]    0    0  720    0  576  288    0
%$
%M = \left(
%\begin{array}{rrrrrrr}
%0 & 0 & 0 & 0 & 0 & 576 & 0\\
%0 & 0 & 720 & 0 & 576 & 288 & 0
%\end{array}
%\right).
%$
%We used the large-sample distribution for $\hat{\beta}$ 
Based on Theorem~\ref{thm:beta-thm}, under $H_0$, 
%Under $H_0$, 
the statistic $(M\hat{\beta})^\intercal \left(M\hat{\Sigma}_\beta M^\intercal \right)^{-1} M\hat{\beta}$,
where $\hat{\Sigma}_\beta$ is the estimated covariance matrix of $\hat{\beta}$,
has a $\chi^2$ distribution with two degrees of freedom in large samples.
This test statistic was equal to $10.32$ with a $p$-value of $0.006$, 
suggesting differences in the AUCs among the three DTRs.
%variation in the 36-week effect of each of these DTRs 
%on the verbal communication skills of children in this study population. 
Following this omnibus test, we examined pairwise contrasts between each DTR
 at each time point, given in Figure~\ref{fig:pairwise-contrasts}, 
 which suggest that the DTR which starts with the AAC speech device
  is superior to the other two DTRs, at least during the first 12 weeks.

\begin{figure}[!p]
\centering
\includegraphics[width=\textwidth]{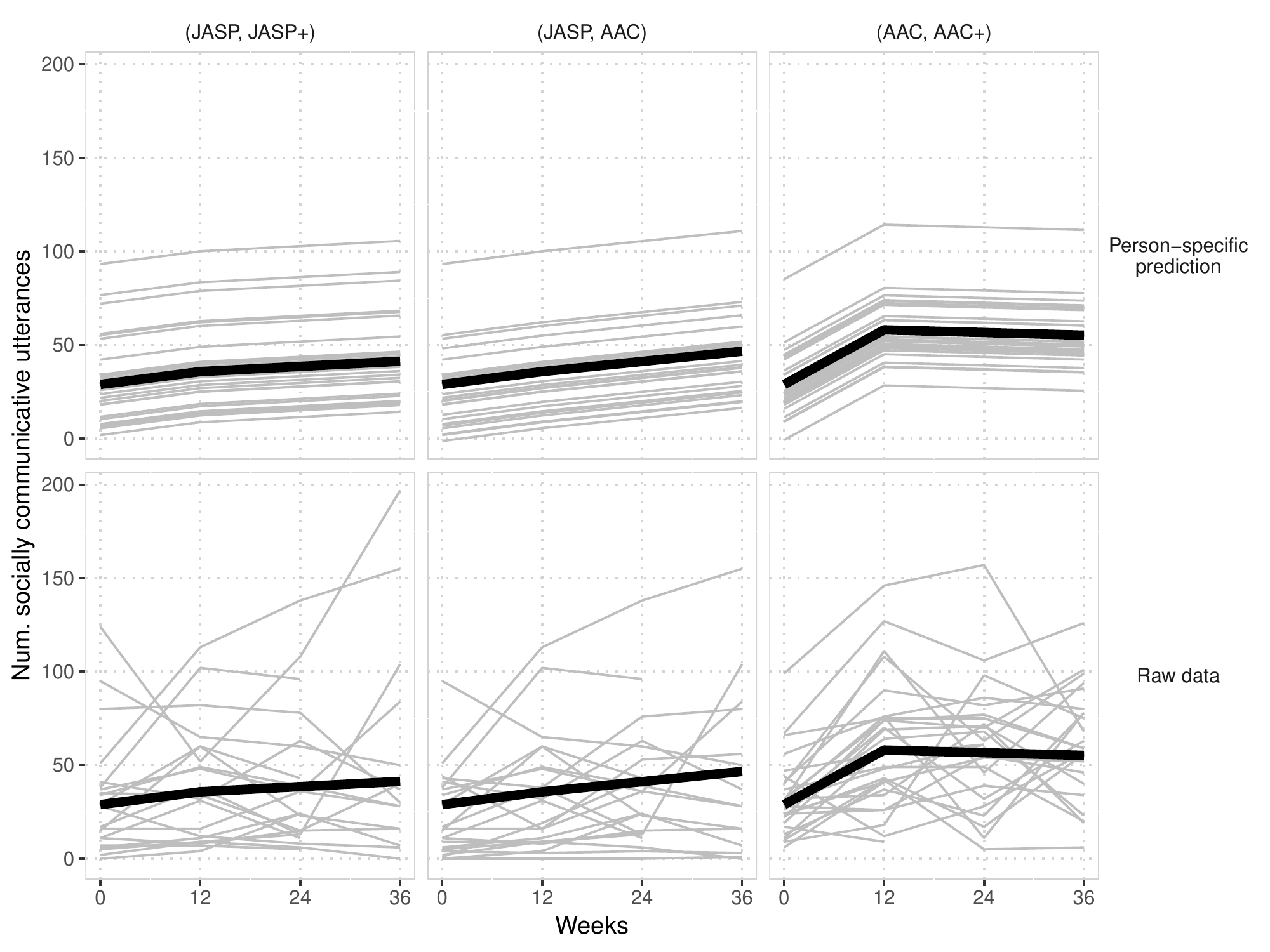}
\caption{Person-specific predicted trajectories using the intercepts-only model for the autism SMART (top row) along with observed number of socially communicative utterances (bottom row). Bold lines are the estimated marginal mean trajectories under each DTR for children at age $6.3$, the average age of the study participants. Responders to initial treatment with JASP are observable under both (JASP, JASP+) and (JASP, AAC), and the observed and predicted trajectories for these participants are displayed for both of these regimens.}
\label{fig:blups}
\end{figure}

%As described in Section~\ref{sec:blup}, a unique feature of mixed models is 
%the ability to predict person-specific trajectories using a plug-in estimator for 
%the posterior mean of the random effects vector, conditional on the observed data. 
For demonstration, Figure~\ref{fig:blups} displays predicted 
person-specific trajectories, $\hat{\beta}^\intercal X_{it}(a_1,a_2) + \hat{b}_i$, 
using the intercepts-only mixed model, 
along with the observed outcomes and
 the estimated mean outcome under each DTR. 
This display could be used to assess subject-to-subject variation relative to 
the estimated mean under each DTR or to identify individuals with 
outlying trajectories based on the fitted model.
In this example, random intercepts lead to
 subject-specific trajectories which
are parallel to the estimated mean under each DTR.
The potential high outliers under the DTRs
(JASP, AAC) and (AAC, AAC+) could be investigated to help characterize 
the variation in communication outcomes for
these study participants.

\section{Discussion}
\label{sec:discussion}
%
%In this manuscript we showed how to use mixed models,
%a standard method for longitudinal data analysis, for primary aim
%comparisons between embedded DTRs in a SMART.

In Section~\ref{sec:blup} we proposed a method for
 predicting random effects based on a weighted pseudo-likelihood.
%used to estimate the parameters in a linear model for the mean. 
The prediction method we propose is analogous to the
``best linear unbiased predictors'' 
commonly used in standard mixed effects analysis of longitudinal data
(\citeauthor{robinson1991}, \citeyear{robinson1991};
\citeauthor{verbeke2009linear}, \citeyear{verbeke2009linear}, Section 7.4). 
However, our proposed predictor $\hat{b}_i$ is a nonlinear
function of $(A_{1i}, R_i A_{2i})$ across all $i$,
 and it is unclear whether
$\hat{b}_i$ has minimum mean squared error (MSE) marginally over
these random variables.
%Even though $\hat{\mathbf{b}}$ can be written as $m^\intercal \mathbf{Y}$,
%a linear combination of $\mathbf{Y}$, the corresponding coefficient vector $m$
Further work is needed to derive a minimum-MSE
property for $\hat{b}_i$ which is marginal over $(A_{1i}, R_i, A_{2i})$ and uses
the same statistical and causal assumptions of Theorem~\ref{thm:beta-thm}. 

%%%%% PARAGRAPH ABOUT SOFTWARE IMPLEMENTATION
%The software implementation %of the weighted pseudo-likelihood
%used for the analysis in Section~\ref{sec:application} 
%is limited to cases where the weights $W_i^{(a_1,a_2)}(R_i)$
%are integers. 
%Additional work is needed for a general implementation of the
%weighted pseudo-likelihood in cases where
%the weights are not integers, which may occur when
%the randomization probabilities are unequal
%across treatment options, or when the weights
%are estimated (e.g. \cite{williamson_EffEstWeightsInRCT:2014, hirano2003efficient}).

%Although we focused on SMARTs with a longitudinal outcome,
% the mixed model developed in this paper could, in principle,
%be used to estimate the end-of-study marginal mean for cluster-level
%DTRs, as in \citet{necamp2017comparing-DTRs-cluster-SMART}, 
%by 
%modifying the marginal mean model $\beta^\intercal X_i(a_1,a_2)$  
%to no longer be a function of time. 
%An exchangeable correlation structure
%for clustered end-of-study outcomes could be modeled with
%a random intercept for each cluster. 
%Another direction for future research is 
%to develop a generalized linear mixed model for 
%SMARTs with a longitudinal binary outcome.
% developed in this manuscript to accommodate longitudinal measurements
% collected in a SMART for cluster-level DTRs.
% This may be possible
%by carefully choosing a random effects structure which 
%models within-subject correlation across time as well as 
%subject-to-subject correlation within a cluster.

This article focused on marginal mean models for the embedded DTRs 
that are conditional only on baseline
covariates. This is analogous to primary aim analyses
in standard randomized trials.
 An alternative approach would be to
specify a mixed model conditional on both the baseline
covariates and the embedded tailoring variables. 
For example, in the autism SMART, one could propose
a mixed effects model for $Y_i$ conditional on $A_{1i}, R_i, A_{2i}$.
%Rather than modeling
%changes in the mean of the potential outcome $Y_i(a_1,a_2)$ 
%had all individuals followed DTR $(a_1,a_2)$, this alternative
%approach would model change in the \emph{observed} outcome over time.
Future work will investigate how to obtain consistent estimators for %for the same
marginal estimands using % of interest in this manuscript using 
this kind of conditional modeling of the observed longitudinal outcome.

In addition to the reasons given in Section \ref{sec:intro}, 
scientists might prefer mixed effects models 
because they may require less restrictive 
assumptions about missing data, at least 
when the true probability distribution
for the observed data is correctly specified 
(\citeauthor{hedeker2006longitudinal}, \citeyear{hedeker2006longitudinal}, Ch. 14;
\citeauthor{fitzmaurice2012applied}, \citeyear{fitzmaurice2012applied}, Ch. 17)
%(\citet[Chapter 14]{hedeker2006longitudinal}, 
%\citet[Chapter 17]{fitzmaurice2012applied}).
%\citet{gibbons2010advances}). 
Our marginal modeling and weighted, pseudo-likelihood
estimation approach does not require
a correct specification of the true probability distribution
that generated the observed data. 
Additional work is needed to understand whether 
our marginal model for longitudinal SMARTs enjoys
the purported benefits of standard mixed models 
in the presence of missing data. 
In the appendix we present simulation results which suggest that our mixed model
may offer some protection against bias in the presence
of ignorable missing data. 
%Finally, as defined in equation~\eqref{eq:mixedmodel},
%our mixed model permits random effects which 
%depend on the treatment regimen $(a_1,a_2)$. 
%Additional work is needed to understand when it is advisable to 
%include DTR-specific random effects terms so that the marginal covariance 
% $V_i(a_1,a_2)$ depends on $(a_1,a_2)$. 
% In some SMART designs the 
%estimate variance parameters for DTR-specific random effects terms,

\section*{Acknowledgments}
This research is supported by the following NIH grants:
		R01DA039901 (Nahum-Shani and Almirall),
		P50DA039838 (Almirall),
		R01HD073975 (Kasari and Almirall), 
                R01MH114203 (Almirall), and 
                R01DA047279 (Almirall).

{\it Conflict of Interest}: None declared.

\begin{table}[!p]
\tblcaption{The dynamic treatment regimens embedded in the example autism SMART.
The last column provides the known inverse probability weight for subjects in each of the cells A--E in Figure~\ref{fig:exampleSMART_autism}.
\label{table:embeddedDTRs}}
{\footnotesize
\begin{tabular}{cccccccc}
		&                      &            & First-stage               & Status at end             & Second-stage                   & Cell in   & Known       \\
	& DTR Label              & $(a_1,a_2)$       &  Treatment        & of first-stage            & Treatment                      & Figure    & IPW    \\ \hline
			&  \multirow{2}{*}{(JASP, JASP+)}      & \multirow{2}{*}{$(1,1)$}  & \multirow{2}{*}{JASP}     & Responder                 & Continue JASP                  & A         & 2      \\
			&                                      &                           &                           & Slow Responder            & Intensify JASP                 & B         & 4      \\ [1.ex]
			&  \multirow{2}{*}{(JASP, AAC)}        & \multirow{2}{*}{$(1,-1)$} & \multirow{2}{*}{JASP}     & Responder                 & Continue JASP                  & A         & 2      \\
			&                                      &                           &                           & Slow Responder            & Augment JASP+AAC               & C         & 4      \\ [1.ex]
			&  \multirow{2}{*}{(AAC, AAC+)}        & \multirow{2}{*}{$(-1,\cdot)$} & \multirow{2}{*}{JASP+AAC} & Responder                 & Continue JASP+AAC              & D         & 2      \\
			&                                      &                           &                           & Slow Responder            & Intensify JASP+AAC             & E         & 2       
\end{tabular}
}
\end{table}

\begin{table}[!p]
\tblcaption{Estimation performance of an end-of-study contrast with two mixed model specifications when the population of potential outcomes exactly follows the marginal distribution implied by the slopes and intercepts mixed model. 
The intercepts only model specifies the correct mean model but is otherwise misspecified. 
Values compuuted from 1,000 simulation replicates. The nominal confidence level was 95 percent.
\label{tab:sim1}}
{\footnotesize
% correct model, LMM slopes and intercepts only
% latex table generated in R 3.5.0 by xtable 1.8-2 package
% Wed Sep 11 11:29:39 2019
\begin{tabular}{lrrrrrrrr}
 Method & $d$ & True value & N & Bias & 
 \begin{tabular}{@{}r@{}}Monte \\ Carlo SD \end{tabular} & 
 \begin{tabular}{@{}r@{}}SE \\ Estimate \end{tabular} & 
 \begin{tabular}{@{}r@{}}CI \\ Coverage \end{tabular} & RMSE \\ 
% Monte Carlo SD & SE estimate & CI Coverage 
  \hline
LMM slopes and intercepts & 0.2 & 0.600 & 50 & -0.102 & 1.222 & 1.119 & 0.911 & 1.226 \\ 
   &  &  & 200 & -0.018 & 0.659 & 0.629 & 0.931 & 0.659 \\ 
   &  &  & 1000 & 0.010 & 0.296 & 0.290 & 0.945 & 0.296 \\ 
   & &  & 5000 & -0.002 & 0.132 & 0.130 & 0.945 & 0.132 \\ 
   & 0.8 & 2.480 & 50 & 0.071 & 1.228 & 1.117 & 0.905 & 1.229 \\ 
   &  &  & 200 & -0.002 & 0.661 & 0.623 & 0.932 & 0.661 \\ 
   &  &  & 1000 & -0.008 & 0.285 & 0.286 & 0.950 & 0.285 \\ 
   &  &  & 5000 & -0.007 & 0.129 & 0.128 & 0.948 & 0.129 \\ 
  LMM intercepts only & 0.2 & 0.600 & 50 & -0.018 & 1.338 & 1.196 & 0.886 & 1.338 \\ 
   &  &  & 200 & 0.001 & 0.748 & 0.694 & 0.917 & 0.748 \\ 
   &  &  & 1000 & 0.010 & 0.336 & 0.323 & 0.938 & 0.336 \\ 
   &  &  & 5000 & 0.002 & 0.146 & 0.145 & 0.951 & 0.146 \\ 
   & 0.8 & 2.480 & 50 & 0.148 & 1.339 & 1.191 & 0.888 & 1.346 \\ 
   & &  & 200 & 0.011 & 0.728 & 0.682 & 0.922 & 0.727 \\ 
   & &  & 1000 & -0.006 & 0.312 & 0.317 & 0.957 & 0.311 \\ 
   & &  & 5000 & -0.005 & 0.143 & 0.142 & 0.957 & 0.143 \\ 
   \hline
\end{tabular}
}
\end{table}

\begin{table}[!p]
\tblcaption{Simulation 2: Estimation of an end-of-study contrast with true value $2.1197$ 
and standardized effect size $d \approx 0.5$. 
Values computed from 1,000 simulation replicates. 
The nominal confidence level was 95 percent.
RMSE inflation is the ratio of the RMSE
to the smallest RMSE among the five methods for a fixed sample size.
\label{tab:sim2}
}
{\footnotesize
% misspecification, compare to GEE
% medium effect size only
% latex table generated in R 3.5.0 by xtable 1.8-2 package
% Mon Oct 14 14:19:30 2019
\begin{tabular}{rlrrrrrr}
% N & Method & Bias & Monte Carlo SD & SE estimate & CI Coverage & RMSE Inflation & $\|V_{\text{true}} - \expected{\hat{V}}\| / \|V_{\text{true}}\|$ \\ 
 N & Method & Bias & \begin{tabular}{@{}r@{}}Monte \\ Carlo SD \end{tabular} & 
 \begin{tabular}{@{}r@{}}SE \\ Estimate \end{tabular} & 
 \begin{tabular}{@{}r@{}}CI \\ Coverage \end{tabular} & 
 \begin{tabular}{@{}r@{}}RMSE \\ Inflation \end{tabular} & $\frac{\|V_{\text{true}} - \expected{\hat{V}}\|}{\|V_{\text{true}}\|}$ \\ 
  \hline
50 & LMM slopes and intercepts & 0.013 & 1.655 & 1.505 & 0.910 & 1.000 & 0.051 \\ 
   & GEE Unstructured & 0.064 & 1.761 & 1.452 & 0.854 & 1.064 & 0.107 \\ 
   & LMM intercepts only & 0.115 & 1.922 & 1.656 & 0.871 & 1.163 & 0.640 \\ 
   & GEE Exchangeable & 0.114 & 1.924 & 1.651 & 0.870 & 1.164 & 0.643 \\ 
   & GEE Independence & 0.182 & 2.163 & 1.804 & 0.861 & 1.311 & 0.900 \\ 
  200 & LMM slopes and intercepts & -0.041 & 0.842 & 0.839 & 0.938 & 1.000 & 0.015 \\ 
   & GEE Unstructured & -0.019 & 0.878 & 0.822 & 0.925 & 1.042 & 0.028 \\ 
   & LMM intercepts only & 0.002 & 0.983 & 0.951 & 0.930 & 1.167 & 0.638 \\ 
   & GEE Exchangeable & 0.002 & 0.984 & 0.950 & 0.931 & 1.167 & 0.638 \\ 
   & GEE Independence & 0.016 & 1.099 & 1.055 & 0.936 & 1.305 & 0.900 \\ 
  1000 & LMM slopes and intercepts & -0.006 & 0.396 & 0.385 & 0.948 & 1.000 & 0.009 \\ 
   & GEE Unstructured & -0.003 & 0.410 & 0.384 & 0.933 & 1.034 & 0.011 \\ 
   & LMM intercepts only & 0.011 & 0.442 & 0.439 & 0.950 & 1.115 & 0.637 \\ 
   & GEE Exchangeable & 0.011 & 0.442 & 0.439 & 0.949 & 1.115 & 0.637 \\ 
   & GEE Independence & 0.021 & 0.483 & 0.489 & 0.950 & 1.220 & 0.900 \\ 
  5000 & LMM slopes and intercepts & 0.000 & 0.172 & 0.174 & 0.958 & 1.000 & 0.007 \\ 
   & GEE Unstructured & -0.001 & 0.178 & 0.174 & 0.947 & 1.039 & 0.007 \\ 
   & LMM intercepts only & 0.005 & 0.204 & 0.198 & 0.936 & 1.191 & 0.637 \\ 
   & GEE Exchangeable & 0.005 & 0.204 & 0.198 & 0.936 & 1.191 & 0.637 \\ 
   & GEE Independence & 0.005 & 0.227 & 0.221 & 0.947 & 1.323 & 0.900 \\ 
   \hline
\end{tabular}
}
\end{table}

% latex table generated in R 3.5.0 by xtable 1.8-2 package
% Wed May 22 11:30:32 2019
\begin{table}[!p]
\tblcaption{Coefficient estimates, standard errors (SE) and 95-percent confidence
intervals from the random intercepts mixed model for the autism SMART.
\label{tab:betahat}
}{
\begin{tabular}{rlll}
Coefficient  & Estimate & SE & 95\% CI \\ 
\hline
 $\beta_0$& 28.885 & 3.763 & $(21.509, 36.261)$ \\ 
  $\beta_1$ & 1.501 & 0.315 & $(0.885, 2.118)$ \\ 
  $\beta_2$ & $-0.929$ & 0.287 & $(-1.492, -0.367)$ \\ 
  $\beta_3$ & 0.112 & 0.174 & $(-0.229, 0.452)$ \\ 
  $\beta_4$ & 0.23 & 0.174 & $(-0.111, 0.571)$ \\ 
  $\beta_5$& $-0.111$ & 0.137 & $(-0.38, 0.158)$ \\ 
  $\beta_6$ & $-4.514$ & 2.777 & $(-9.957, 0.93)$ \\ 
\end{tabular}
}
\end{table}

\newpage

\clearpage 

\appendix

% \section{Appendices}
This appendix contains the following:
a proof of Theorem~\ref{thm:beta-thm}; 
further description of the GEE estimator
used as a comparison method in the simulation studies 
in Section~\ref{sec:sims};
results of a simulation study with ignorable missing data; 
and a description
of how the weighted pseudo-likelihood
was implemented with standard mixed models software
for the analysis in Section~\ref{sec:application}.

\section{Proof of Theorem 4.1}
\label{app:beta-thm-proof}
First note that $\sum_iU_i(\hat\beta, \hat\alpha) = 0$, $R_i(1-R_i)=0$, 
$R_i^2=R_i$ and $(1-R_i)^2=1-R_i$. To simplify notation, 
we suppress dependence of $X_i(a_1,a_2)$ and 
$V_i(a_1,a_2; \alpha^*)$ on $(a_1,a_2)$. Then, by definition of $\tilde{W}_i(a_1,a_2)$,

	\begin{align}
\begin{split}
	&\expected{\tilde{W}_i(a_1,a_2) X_i^\intercal V_i( \alpha^*)^{-1} (Y_i-X_i\beta^*) \given L_i}\\
	&=\mathbb{E}\left[\frac{\indic{A_{1i}=a_1}}{\prob{A_{1i}=a_1}}\left(R_i + \frac{\indic{A_{2i}=a_2}}{\prob{A_{2i} = a_2 \given A_{1i}=a_1, R_i=0}}(1-R_i)\right) \right. \\
	&\hspace{48pt} \left. \vphantom{[\frac{\indic{A_{1i}=a_1}}{\prob{A_{1i}=a_1}}} \times X_i ^\intercal V_i( \alpha^*)^{-1}(Y_i-X_i\beta^*) \given L_i\right],
\end{split}
	\end{align}
and, using consistency assumption (ii)
	\begin{align}
	\begin{split}
	&=\expected{\frac{\indic{A_{1i}=a_1}}{\prob{A_{1i}=a_1}} R_i X_i^\intercal V_i(\alpha^*)^{-1}(R_iY_i(A_{1i})-X_i \beta^*)\given L_i}\\[6pt]
	&\hspace{12pt} + \mathbb{E}\left[\frac{\indic{A_{1i}=a_1}}{\prob{A_{1i}=a_1}}\frac{\indic{A_{2i}=a_2}}{\prob{A_{2i} = a_2|A_{1i}=a_1, R_i=0}}(1-R_i) \right.\\
	&\hspace{48pt} \left.  \vphantom{\frac{\indic{A_{1i}=a_1}}{\prob{A_{1i}=a_1}}} \times X_i^\intercal V_i( \alpha^*)^{-1} (Y_i(A_{1i}, A_{2i})-X_i\beta^*)\given  L_i\right].
	\end{split} \label{eq:consistency-step}
	\end{align}
Next note that $\indic{A_{2i}=a_2}Y_i(A_{1i},A_{2i})=\indic{A_{2i}=a_2}Y_i(A_{1i},a_2)$ 
and, by assumption (iii), 
$A_{2i} \indep Y(a_1,a_2) \given A_{1i},R_i$
 for any fixed regime $(a_1,a_2)$. 
  Let $Q = \prob{A_{1i}=a_1}^{-1}\prob{A_{2i}=a_2 \given A_{1i}=a_1,R_i=0}^{-1}$.
	\begin{align}
&\expected{Q \indic{A_{1i}=a_1} \indic{A_{2i}=a_2} (1-R_i)  X_i^\intercal V_i( \alpha^*)^{-1} (Y_i(A_{1i}, A_{2i})-X_i\beta^*)\given  L_i}\\
&=\mathbb{E}\left\{ Q \indic{A_{1i}=a_1}\expected{  \indic{A_{2i}=a_2} (1-R_i)  X_i^\intercal V_i( \alpha^*)^{-1} (Y_i(A_{1i}, A_{2i})-X_i\beta^*)\given A_{1i}, R_i} \given L_i \right\}\\
&=\mathbb{E}\left\{Q\indic{A_{1i}=a_1}\expected{  \indic{A_{2i}=a_2} (1-R_i)  X_i^\intercal V_i( \alpha^*)^{-1} (Y_i(A_{1i}, a_2)-X_i\beta^*)\given A_{1i}, R_i =0} \given L_i \right\}\\
%&=\mathrm{E}\left\{Q\indic{A_{1i}=a_1}\prob{A_{2i}=a_2\given A_{1i}=a_2,R_i=0} \expected{ (1-R_i)  X_i^\intercal V_i( \alpha^*)^{-1} (Y_i(A_{1i}, a_2)-X_i\beta^*)\given A_{1i}, R_i =0} \given L_i \right\}\\
&=\mathrm{E}\left\{\frac{\indic{A_{1i}=a_1}}{\prob{A_{1i}=a_1}}\expected{ (1-R_i)  X_i^\intercal V_i( \alpha^*)^{-1} (Y_i(A_{1i}, a_2)-X_i\beta^*)\given A_{1i}, R_i =0} \given L_i \right\}
	\end{align}
Substituting into equation~\eqref{eq:consistency-step},  
	\begin{align}
	\begin{split}
\text{\eqref{eq:consistency-step}}	= & \expected{\frac{\indic{A_{1i}=a_1}}{\prob{A_{1i}=a_1}} R_i X_i^\intercal V_i( \alpha^*)^{-1}(R_iY_i(A_{1i})-X_i\beta^*) \given  L_i}\\
	&+ \expected{\frac{\indic{A_{1i}=a_1}}{\prob{A_{1i}=a_1}}(1-R_i) X_i^\intercal V_i( \alpha^*)^{-1}(Y_i(A_{1i}, a_{2})-X_i \beta^*) \given L_i}
	\end{split}\\
	= & X_i^\intercal V_i( \alpha^*)^{-1}
	\expected{R_i(a_1)Y_i(a_{1})+(1-R_i(a_1))Y_i(a_1, a_2)-X_i\beta^* \given L_i} \label{eq:penultimate}\\
	= & 0
	\end{align}
where \eqref{eq:penultimate} is obtained from the consistency assumption on $R_i$ and independence of $A_{1i}$ and $R(a_1)$.
	Thus $\expected{\sum_iU_i(\beta^*,\alpha^*)}=0$. Under Assumption (v), we have that $\hat{\beta}$ is a consistent estimator of $\beta^*$. To derive the asymptotic distribution of $\hat{\beta}$, note that
	\begin{align*}
	\sqrt N (\hat\beta-\beta^*) = - \left(\frac{1}{N}\frac{d \sum_iU_i(\beta, \hat\alpha (\beta))}{d\beta} \Big|_{\beta=\beta^*} + o_P(1)\right)^{-1} \frac{1}{\sqrt N}\sum_iU_i(\beta^*, \hat\alpha (\beta^*))
	\end{align*}
	The result follows using similar arguments as those in the proof of Theorem 2 in \citet{liang1986longitudinal}.

\section{Details of the GEE estimator used in simulation}
Here we describe the GEE-like estimator of
\citet{lu:2016} and \citet{seewald2018-repeatedmeasures-power-compareDTRs}
used as a comparison method in the simulations in Section 5.

First, an initial least squares estimate is computed:
$$\hat{\beta}^{(0)} =  \left(\sum_i\sum_{a_1,a_2}\tilde{W}_i(a_1,a_2)X_i(a_1,a_2)^\intercal X_i(a_1,a_2)\right)^{-1}\left(\sum_i\sum_{a_1,a_2}\tilde{W}_i(a_1,a_2)X_i(a_1,a_2)^\intercal Y_i\right).$$ 
This initial estimate is used to compute the residual vectors
$r_i^{(0)}(a_1,a_2) = Y_i-X_i(a_1,a_2)\hat{\beta}^{(0)}$ for all $i$ and 
$(a_1,a_2)$. 

Next we compute method of moments estimators for 
$V_i(a_1,a_2)$. Let $D$ be the number of embedded DTRs, i.e. $D = \sum_{a_1,a_2} 1$.
For $t \neq s$, define the following:
\begin{align*}
\hat{\sigma}_t^2(a_1,a_2)&=\frac{\sum_i\tilde{W}_i(a_1,a_2)r_{it}^{(0)}(a_1,a_2)^2}{N_t} & \hat{\sigma}^2(a_1,a_2)&=\frac{N_t\hat{\sigma}_t^2(a_1,a_2)}{\sum_t N_t}\\
\hat{\sigma}^2_t &= \frac{1}{D}\sum_{a_1,a_2} \hat{\sigma}^2_t(a_1,a_2) & \hat{\sigma}^2 &= \frac{1}{D}\sum_{a_1,a_2} \hat{\sigma}^2(a_1,a_2)\\
\hat{\rho}_{ts}(a_1,a_2) &= \frac{1}{N_{ts}} \sum_i \tilde{W}_i(a_1,a_2)\left[\frac{r_{it}^{(0)}(a_1,a_2) r_{is}^{(0)}(a_1,a_2)}{\hat{\sigma}_t(a_1,a_2)\hat{\sigma}_s(a_1,a_2)}\right], & \hat{\rho}_{ts} &= \frac{1}{D}\sum_{a_1,a_2}\hat{\rho}_{ts}(a_1,a_2)\\
\hat{\rho}(a_1,a_2)&=\frac{1}{N}\sum_i\frac{\tilde{W}_i(a_1,a_2)}{n_i(n_i-1)/2} \sum_{s<t}\frac{r_{is}^{(0)}(a_1,a_2)r_{it}^{(0)}(a_1,a_2)}{\hat{\sigma}_s(a_1,a_2)\hat{\sigma}_t(a_1,a_2)}, &\hat{\rho} &= \frac{1}{D}\sum_{a_1,a_2}\hat{\rho}(a_1,a_2)\\[6pt]
\hat{\psi}(a_1,a_2)&=\frac{1}{N}\sum_i \frac{\tilde{W}_i(a_1,a_2)}{n_i(n_i-1)/2}\sum_{s<t} \frac{r_{is}^{(0)}(a_1,a_2)r_{it}^{(0)}(a_1,a_2)}{\hat{\sigma}^2(a_1,a_2)},&\hat{\psi} &= \frac{1}{D}\sum_{a_1,a_2}\hat{\psi}(a_1,a_2)
\end{align*}
where $N_t$ is the number of individuals with an observation 
at unique time point $t$ and $N_{ts}$ is the number of 
individuals with observations at both of the time points $t$ and $s$. 
The estimators defined above are simply the method of moments estimators for correlation or variance parameters at each observation. 
They differ in whether the variances or correlations are assumed
to be equal across DTRs and in whether the variance is
assumed to be constant as a function of time.
By combining these correlation and variance estimators, we can obtain 
various working models for $V_i(a_1,a_2)$. 
For example, 
the unstructured and exchangeable estimates of $V_i(a_1,a_2)$ have the following entries, $v_{ts}(a_1,a_2)$: 
\begin{align*}
\text{Unstructured}\hspace{12pt}& v_{tt}(a_1,a_2)=\hat{\sigma}^2_t & \text{Exchangeable}\hspace{12pt}&v_{tt}(a_1,a_2)=\hat{\sigma}^2 \text{ for all } t\\
& v_{ts}(a_1,a_2)=\hat{\sigma}_s\hat{\sigma}_t\hat{\rho}_{ts}  & &v_{ts}(a_1,a_2) = \hat{\psi}\hat{\sigma}^2\\
 & \text{ for all } a_1, a_2 & & \text{for all } a_1,a_2
\end{align*}
The independence working model sets all off-diagonal entries of $V_i(a_1,a_2)$ to zero and all diagonal entries to $\hat{\sigma}^2$.

Autoregressive working models for $V_i(a_1,a_2)$ are also possible using the following correlation estimators:
\begin{align*}
\hat{\tau}_t(a_1,a_2) &= \frac{1}{N}\sum_i \frac{\tilde{W}_i(a_1,a_2)}{n_i-1}\sum_{s=1}^{n_i-1}\frac{r_{is}^{(0)}(a_1,a_2)r_{i(s+1)}^{(0)}(a_1,a_2)}{\hat{\sigma}_s(a_1,a_2)\hat{\sigma}_{s+1}(a_1,a_2)}\\
\hat{\tau}_t &= \frac{1}{D}\sum_{a_1,a_2}\hat{\tau}_t(a_1,a_2)\\
 \hat{\tau}(a_1,a_2)&=\frac{1}{N}\sum_i\frac{\tilde{W}_i(a_1,a_2)}{n_i-1}\sum_{s=1}^{n_i-1}\frac{r_{is}^{(0)}(a_1,a_2) r_{i(s+1)}^{(0)}(a_1,a_2)}{\hat{\sigma}^2(a_1,a_2)} \\
 \hat{\tau} &= \frac{1}{D}\sum_{a_1,a_2}\hat{\tau}(a_1,a_2)%\frac{1}{DN}\sum_{a_1,a_2}\sum_i \frac{\tilde{W}_i(a_1,a_2)}{n_i-1}\sum_{s=1}^{n_i-1}\frac{r_{is}r_{i(s+1)}}{\hat{\sigma}^2}
\end{align*}

\section{Details of the generative model used in simulation}
This section provides more detail about the generative model
used in the simulation studies in Section 5.

The potential outcomes were
generated from the following 
 piecewise linear model: 
    \begin{align}
    \begin{split}
    Y_{it}(a_1,a_2) &= \\
    \hspace{12pt}&\theta_0 + \indic{t \le \kappa} t(\theta_1+\theta_2 a_1) + \indic{t > \kappa} \kappa (\theta_1+\theta_2 a_1)\\
    	&+ \indic{t > \kappa}(t-\kappa)(\theta_3+\theta_4 a_1 + (\theta_5 a_2+\theta_6 a_1a_2)(1-R_i(a_1)))\\
    	&+ \indic{t > \kappa}(t-\kappa)(\psi^{(1)}\indic{a_1=1} + \psi^{(-1)}\indic{a_1=-1})\left[R_i(a_1)-\prob{R_i(a_1)=1\given L_i}\right]\\
    	&+\theta_7 L_i + \gamma_{0i} + \gamma_{1i} t + \epsilon_{it},
    \end{split}\label{eq:sim-gen}
    \end{align}
where $R_i(a_1) = \indic{Y_{i \kappa}(a_1) -\theta_7 L_i > c}$; 
$c = 1.1$; $\theta_7 = -0.2$;
$ (\gamma_{0i}, \gamma_{1i})^\intercal \sim N(0,\Gamma)$;
$\epsilon_{it} \sim N(0, \tau^2)$ with $\tau^2 = 1$;
$ t \in \left\{0, 0.5, 1.5, 2, 2.25, 2.5, 3\right\}$; and $\kappa = 2$.
%In this generative model, 
%$R_i(a_1)$ is a function of the final first-stage outcome $Y_{i\kappa}(a_1)$.
%In each sample of $N$ simulated participants, half of the participants were assigned $L_i = 1$ and the other half were assigned $L_i=-1$. 
%The parameters $\psi^{(1)},\psi^{(-1)}$ induce a marginal association 
%between response status and second-stage outcomes, 
%and the fixed value of $c$ in this generative model means that 
%$\pi^{(a_1)} := \prob{R_i(a_1) = 0\given L_i}$ 
% varies as a function of $a_1$. 
  
Under model \eqref{eq:sim-gen},
%\begin{align}
%\begin{split}
%&\expected{Y_{it}(a_1,a_2)\given L_i}\\
%&\hspace{12pt} = \theta_0 + \indic{t \le \kappa } t (\theta_1+\theta_2 a_1) + \indic{t > \kappa} \kappa (\theta_1+\theta_2 a_1)\\
%&\hspace{18pt}+\indic{t > \kappa}(t-\kappa)\left(\theta_3+\theta_4a_1+\theta_5\prob{R_i(a_1)=0\given X} a_2 + \theta_6\prob{R_i(a_1) = 0 \given L_i} a_1a_2\right)
%\end{split}
%\end{align}
%and 
\begin{align}
\begin{split}
&Y_{it}(a_1,a_2) - \expected{Y_{it}(a_1,a_2) \given L_i}\\
&\hspace{12pt}= \gamma_0 + \gamma_1 t + \epsilon_t\\
&\hspace{12pt}+\indic{t > \kappa}(t-\kappa)\theta_5 a_2\left[(1-R_i(a_1)) - \prob{R_i(a_1)=0\given L_i}\right]\\
&\hspace{12pt}+\indic{t > \kappa}(t-\kappa)\theta_6 a_1a_2\left[(1-R_i(a_1))-\prob{R_i(a_1)=0 \given L_i}\right]\\
&\hspace{12pt}+\indic{t > \kappa}(t-\kappa)\left(\psi^{(1)}\indic{a_1=1}+\psi^{(-1)}\indic{a_1=-1}\right)\left[R_i(a_1)-\prob{R_i(a_1)=1 \given L_i}\right],
\end{split}
\end{align}
and we can parameterize this marginal mean model as follows:
%model can be written as follows:
  \begin{align}
    \begin{split}
    \expected{Y_{it}(a_1,a_2) \given L_i} = 
    \beta^\intercal X_{it}(a_1,a_2) = \beta_0&+\indic{t \le \kappa}t (\beta_1+\beta_2 a_1) + \indic{t > \kappa}\kappa(\beta_1+\beta_2 a_1)\\
    	&+\indic{t > \kappa}(t-\kappa)(\beta_3+\beta_4 a_1+\beta_5 a_2 + \beta_6 a_1a_2)\\
	    &+ \beta_7 L_i,
    \end{split}\label{eq:simmean}
    \end{align}
    where $\beta_j = \theta_j$ for $j \in \left\{0, 1, 2, 3, 4, 7\right\}$ and
\begin{align*}
\beta_5 &=\left\{\theta_5\left(\frac{\pi^{(1)}}{2}+\frac{\pi^{(-1)}}{2}\right) + \theta_6\left(\frac{\pi^{(1)}}{2}-\frac{\pi^{(-1)}}{2}\right)\right\} \\
\beta_6 &= \left\{ \theta_5\left(\frac{\pi^{(1)}}{2}-\frac{\pi^{(-1)}}{2}\right) + \theta_6 \left(\frac{\pi^{(1)}}{2}+\frac{\pi^{(-1)}}{2}\right)    \right\}. 
\end{align*} 
%\begin{align*}
%\beta_0 &= \theta_0 & \beta_1 &= \theta_1\\
%\beta_2 &= \theta_2 & \beta_3 &= \theta_3\\
%\beta_4 &= \theta_4 & \beta_7&=\theta_7\\
%\beta_5 &=\left\{\theta_5\left(\frac{\pi^{(1)}}{2}+\frac{\pi^{(-1)}}{2}\right) + \theta_6\left(\frac{\pi^{(1)}}{2}-\frac{\pi^{(-1)}}{2}\right)\right\} &&\\
%\beta_6 &= \left\{ \theta_5\left(\frac{\pi^{(1)}}{2}-\frac{\pi^{(-1)}}{2}\right) + \theta_6 \left(\frac{\pi^{(1)}}{2}+\frac{\pi^{(-1)}}{2}\right)    \right\}
%\end{align*}
 
Next, we derive the marginal covariance and variance of the 
repeated measures outcomes under this generative model. 
These marginal covariances and variances are used to calculate 
the population standardized effect size
$$d = \frac{\expected{Y_{i 3}(1,-1) \given L_i} - \expected{Y_{i 3}(-1,-1)\given L_i} }{\sqrt{\frac{1}{2}\mathrm{Var}\left(Y_{i3}(1,-1)\given L_i \right) + \frac{1}{2} \mathrm{Var}\left(Y_{i3}(-1,-1) \given L_i\right)}}.$$

Let
 $W_{it} = z_{it}^\intercal \gamma + \epsilon_{it}$ and $z_{it} = (1, t)^\intercal $.
% and $\pi^{(a_1)}=\prob{R_i(a_1)=0 \given L_i}$. 
 Then
\begin{align}
\begin{split}
&\mathrm{Cov}\left[Y_{it}(a_1,a_2), Y_{is}(a_1,a_2) \given L_i \right] \\
&= z_{it}^\intercal G z_{is} + \tau^2 \indic{s=t} \\
&\hspace{12pt} - \left(C_1^{(a_1)}(s)+C_2^{(a_1,a_2)}(s)-C_3^{(a_1)}(s)\right)(1- \pi^{(a_1)})\expected{W_{it} \given R_i(a_1)=1, L_i}\\
&\hspace{12pt}-\left(C_1^{(a_1)}(t)+C_2^{(a_1,a_2)}(t)-C_3^{(a_1)}(t)\right)(1 - \pi^{(a_1)})\expected{W_{is} \given R_i(a_1)=1, L_i}\\
&\hspace{12pt}+ \left(C_1^{(a_1)}(t)+C_2^{(a_1,a_2)}(t)-C_3^{(a_1)}(t)\right)\left(C_1^{(a_1)}(s)+C_2^{(a_1,a_2)}(s)-C_3^{(a_1)}(s)\right)\pi^{(a_1)}(1-\pi^{(a_1)})
\end{split}
\end{align}
where
\begin{align*}
C_1^{(a_2)}(t) &= \indic{t > \kappa} (t-\kappa)\theta_5 a_2\\
C_2^{(a_1,a_2)}(t)&=\indic{t > \kappa}(t-\kappa)\theta_6 a_1a_2\\
C_3^{(a_1)}(t)&=\indic{t > \kappa}(t-\kappa) (\psi^{(1)}\indic{a_1=1}+\psi^{(-1)}\indic{a_1=-1}).
\end{align*}

Note that 
\begin{align*}
\expected{W_{it} \given R_i(a_1)=1, L_i} &= \expected{W_{it} \given W_{i 
\kappa} > c - \theta_0 -\kappa(\theta_1+\theta_2 a_1), L_i},\\[6pt]
(W_{it}, W_{i \kappa})^\intercal \given L_i &\sim N\left(\left[\begin{array}{c}
0\\0\end{array}\right],\left[
\begin{array}{cc}
z_{it}^\intercal G z_{it} +\tau^2,  &  z_{it}^\intercal G z_{i \kappa} + \mathrm{Cov}(\epsilon_{it},\epsilon_{i \kappa} \given L_i ) \\
\cdots & z_{i \kappa}^\intercal G z_{i \kappa}+\tau^2
\end{array}\right]\right),
\end{align*}
and since $(W_{it}, W_{i \kappa})^\intercal \given L_i$ is bivariate Gaussian, $\expected{W_{it} \given W_{i \kappa} > c-\theta_0-\kappa(\theta_1+\theta_2 a_1), L_i}$ can be computed 
using the truncated multivariate Gaussian distribution.

\section{Additional simulation study}
% missing_cutoff <- -3.5
% missing_timepoint <- 2.25
%as_tibble(dobs) %>%
%  filter(time == missing_timepoint) %>%
%  mutate(ylow = Y < missing_cutoff) %>%
%  group_by(A1,A2,R) %>%
%  summarise(prop_miss = sum(ylow) / n())
%    # A tibble: 6 x 4
%    # Groups:   A1, A2 [?]
%         A1    A2     R prop_miss
%      <dbl> <dbl> <dbl>     <dbl>
%    1    -1    -1     0    0.196 
%    2    -1     1     0    0.171 
%    3    -1    NA     1    0     
%    4     1    -1     0    0.110 
%    5     1     1     0    0.0897
%    6     1    NA     1    0     
One potential benefit of mixed models is their ability
to provide unbiased parameter estimates
when data are missing at random,
assuming that estimation and inference are based on
a correctly specified likelihood for the observed data
(\citeauthor{fitzmaurice2012applied}, \citeyear{fitzmaurice2012applied}, Ch. 17;
\citeauthor{hedeker2006longitudinal}, \citeyear{hedeker2006longitudinal}, Ch. 14;
\citeauthor{molenberghs2007missing}, \citeyear{molenberghs2007missing}, Ch. 7;
\citeauthor{gibbons2010advances}, \citeyear{gibbons2010advances}). 
%(\citet[Chapter 14]{hedeker2006longitudinal}, 
%\citet[Chapter 17]{fitzmaurice2012applied}, 
%\citet{gibbons2010advances}
%\citet[Chapter 7]{molenberghs2007missing}).
In the case of the mixed model we propose
for longitudinal SMARTs, estimation and inference
are based on a weighted pseudo-likelihood,
not the true likelihood for the observed data,
so it is not clear whether bias can be avoided
with ignorable missing data in a SMART.

To help understand whether mixed models
provide any protection against bias due to missing data
in a longitudinal SMART,
this additional simulation study 
describes the performance of our mixed model and
the GEE-like estimators of
 \citet{seewald2018-repeatedmeasures-power-compareDTRs} and \citet{lu:2016} 
when data are missing at random (ignorable) due to study dropout.
 In this scenario, if a participant's observed 
 $Y_{it}$ at $t= 2.25$ was less than $-3.5$,
  then all observations from that participant at time points
   $t \ge 2.5$ were discarded. This results in about
    20 percent dropout among participants with 
    $(A_{1i}, A_{2i}, R_i) = (-1,-1,0)$; about 17 percent 
    dropout among participants with $(A_{1i},A_{2i},R_i) = (-1,1,0)$; 
    about nine and ten percent dropout among participants 
    with $(A_{1i},A_{2i},R_i) = (1,1,0)$ and $(1,-1,0)$, respectively;
     and less than $0.1$ percent dropout among participants with $R_i=1$. 
     
Table~\ref{tab:sim3} compares the same estimators from Simulation 2 
in the Section 5 in their ability to estimate 
an end-of-study contrast with effect size $d \approx 0.5$ 
in the presence of the dropout process described above.
In this scenario, we see bias in large samples, although
the degree of bias decreases as 
$\frac{\|V_{\text{true}} - \expected{\hat{V}}\|}{\|V_{\text{true}}\|}$ decreases.
This suggests that the ability of a mixed model to
flexibly model $V_i(a_1,a_2)$, and to 
efficiently estimate $V_i(a_1,a_2)$, may provide some protection against
bias due to ignorable missing data. Since we are not able
to fit a mixed model using the true likelihood for the potential outcomes,
the purported benefits of mixed models in the presence of ignorable missing data
(compared to GEE approaches) might not exist when analyzing longitudinal SMARTs.

\section{Software implementation with integer-valued weights}
%Augmented data ``trick'' for weighted-replicated objective function
Next we describe how the mixed model for longitudinal SMARTs
can be implemented using standard mixed model software, such as
\citet{lme4-bates}. This implementation is limited
to analyses in which the weights  $W_i^{(a_1,a_2)}(R_i)$ are
integer-valued. 
When estimating these probability of treatment weights \citep{hirano2003efficient, brumback2009note}
or when randomization probabilities are unequal across treatment options,
the weights may not be integer-valued. 
Future work will develop software implementations 
for use in SMART designs beyond the special case of integer-valued weights.

Recall that $I^{(a_1,a_2)}(A_{1i},R_i,A_{2i})$ is an indicator 
of whether subject $i$ is observable under regimen $(a_1,a_2)$. 
For example, in the autism study,
$
I^{(a_1,a_2)}(A_{1i},R_i,A_{2i}) = \indic{A_{1i}=a_1}(R_i+(1-R_i)\indic{A_{2i}=a_2}) 
$. 
Let $f(a_1,a_2,Y_i,L_i)$ be an arbitrary function 
of the observed response vector $Y_i$, 
baseline covariates $L_i$, and the DTR $(a_1,a_2)$. 
In the autism SMART and other common designs, responders ($R_i=1$) are 
observable under both of the DTRs $(A_{1i}, 1), (A_{1i}, -1)$. In this case,
\begin{align}
\begin{split}
&\sum_{i=1}^N \sum_{a_1,a_2} I^{(a_1,a_2)}(A_{1i},R_i,A_{2i}) 
W_i^{(a_{1}, a_{2})}(R_i) f(a_1,a_2,Y_i,L_i)\\[6pt]
&= \sum_{i : R_i = 1} 
W_i^{(A_{1i}, 1)}(1)
f(A_{1i}, 1, Y_i, L_i) + 
\sum_{i : R_i=1}
W_i^{(A_{1i},-1)}(1)
f(A_{1i},-1,Y_i,L_i)\\
&\hspace{12pt} + 
\sum_{i : R_i = 0} 
W_i^{(A_{1i}, A_{2i})}(0)
%W_i(A_{1i}, R=0, A_{2i})
f(A_{1i},A_{2i},Y_i,L_i), 
\end{split}
\end{align}
since when $R_i=0$, subject $i$ is observable under DTR $(A_{1i}, A_{2i})$ only.

%For simplicity, suppose that the model for $V_i$ 
%implied by the mixed model does not depend on $(a_1,a_2)$. 
The weighted pseudo-likelihood is
\begin{align}
\begin{split}
l(\beta,\alpha)&=-\frac{1}{2}\sum_{i=1}^N
\sum_{a_1,a_2} \tilde{W}_i(a_1,a_2)\log\det\left[V_i(a_1,a_2)\right] \\
&\hspace{12pt} -
	\frac{1}{2}\sum_{i}\sum_{a_1,a_2} 
			\tilde{W}_i(a_1,a_2) r_i(a_1,a_2)^\intercal V_i(a_1,a_2)^{-1} r_i(a_1,a_2)
			\end{split}\\
\begin{split}
&=-\frac{1}{2}\sum_{i : R_i=1} 
W_i^{(A_{1i}, 1)}(1)
%W_i(A_{1i},R=1,A_2=1)
		\log\det\left[V_i(A_{1i}, 1)\right] \\
&\hspace{12pt}-\frac{1}{2}\sum_{i : R_i=1} 
W_i^{(A_{1i}, -1)}(1)
%	W_i(A_{1i},R=1,A_2=-1)
	\log\det\left[V_i(A_{1i},-1)\right]  \\
&\hspace{12pt}-\frac{1}{2}\sum_{i : R_i=0} 
W_i^{(A_{1i}, A_{2i})}(0)
%W_i(A_{1i},R=0,A_{2i})
\log\det\left[V_i(A_{1i},A_{2i})\right]\\
&\hspace{12pt}-\frac{1}{2}\sum_{i : R_i=1} 
W_i^{(A_{1i}, 1)}(1)
%	W_i(A_{1i},R=1,A_2=1) 
	(Y_i-X_i(A_{1i},1)\beta)^\intercal 
			V_i(A_{1i}, 1)^{-1}(Y_i-X_i(A_{1i},1)\beta)\\
&\hspace{12pt}-\frac{1}{2}\sum_{i : R_i=1} 	
W_i^{(A_{1i}, -1)}(1)
%W_i(A_{1i},R=1,A_2=-1) 
(Y_i-X_i(A_{1i}, -1)\beta)^\intercal 
		V_i(A_{1i}, -1)^{-1}(Y_i-X_i(A_{1i}, -1)\beta)\\
&\hspace{12pt}-\frac{1}{2}
	\sum_{i : R_i=0} 
	W_i^{(A_{1i}, A_{2i})}(0)
%	W_i(A_{1i},R=0,A_{2i}) 
		(Y_i-X_i^{(A_{1i},A_{2i})}\beta)^\intercal 
	V_i(A_{1i}, A_{2i})^{-1}(Y_i-X_i^{(A_{1i},A_{2i})}\beta)
	\end{split} \label{eq:weighted-lik-allterms}
\end{align}

This objective function is equivalent to the 
log-likelihood in a linear mixed effects model 
based on an ``augmented'' data set constructed in the 
following manner. 
For all subjects $i$ whose observed data are observable under
 more than one DTR, duplicate the 
baseline covariates $L_i$ and 
response vectors $Y_i$ once for each of those DTRs.
In the autism SMART, 
subjects with $R_i=1$ are observable under 
$(A_{1i},1)$ and $(A_{2i},-1)$, so the baseline 
covariates and response vectors are duplicated twice.
The design matrices 
$X_i(a_1,a_2)$ and $Z_i(a_1,a_2)$ for
 each replicate are formed by plugging in 
 the values of $(a_1,a_2)$ corresponding to the 
 DTR under which that replicate is observable. 
 The weights for these replicates are formed similarly. 
 Thus, for a subject with $R_i=1$, the augmented data consist of 
\begin{align*}
\left\{X_i(A_{1i},1), Z_i(A_{1i},1) L_i , Y_i, W_i^{(A_{1i}, 1)}(1)
%W_i(A_{1i},R=1,A_2=1)
\right\}\\
\left\{X_i(A_{1i}, -1), Z_i(A_{1i},-1), L_i , Y_i , W_i^{(A_{1i}, -1)}(1)
%W_i(A_{1i},R=1,A_2=-1)
\right\}
\end{align*}
For subjects whose observed data are observable only
under the DTR $(A_{1i}, A_{2i})$, their observed data are 
unchanged and included in the augmented data set.

Indexing the artificial ``subjects'' in the augmented data set by $s=1,\ldots, M$, we have, based on \eqref{eq:weighted-lik-allterms}, 
\begin{align}
l(\beta,\alpha)% &=
%-\frac{1}{2}\sum_{i=1}^N\sum_{a_1,a_2} \tilde{W}_i(a_1,a_2)\log\det\left[V_i(a_1,a_2)\right] -\frac{1}{2}\sum_{i}\sum_{a_1,a_2} \tilde{W}_i(a_1,a_2) r_i^\intercal V_i(a_1,a_2)^{-1} r_i\\
&=-\frac{1}{2}\sum_{s=1}^M W_s \log\det\left[V_s\right] - \frac{1}{2}\sum_{s=1}^M W_s(Y_s-X_s\beta)^\intercal V_s^{-1}(Y_s-X_s\beta),\label{eq:weighted-aug-loglik}
\end{align}
where $W_s, X_s$ and $V_s$ are the values of 
$W_i^{(A_{1i}, A_{2i})}(R_i)$,
%$W_i(A_{1i},R_i,A_{2i})$, 
$X_i(a_1,a_2)$ and $V_i(a_1,a_2)$ evaluated under the DTR
corresponding to replicate $s$ in the augmented data. 
Thus, to find maximum likelihood estimates of the parameters 
$\beta$ and $\alpha$, we can use any software package which maximizes
a weighted log-likelihood of the form \eqref{eq:weighted-aug-loglik}. 
In particular, when $W_s$ is an integer, we can maximize
\eqref{eq:weighted-aug-loglik} by duplicating all of the terms
indexed by $s$ a total of $W_s$ times and fitting the mixed model
corresponding to \eqref{eq:weighted-aug-loglik} without the use
of weights.

%\newpage

%\bibliographystyle{biorefs}
%\bibliography{refs}

\begin{table}[!p]
\tblcaption{Simulation 3: Estimation performance with ignorable missing data due to study dropout. 
The true contrast value is $2.1197$, corresponding to a standardized effect size
of $d \approx 0.5$. Values computed from 1,000 simulation replicates.
The nominal confidence level was 95 percent.
\label{tab:sim3}
}
{\footnotesize
% dropout simulation 
% medium effect size only
% dropout simulation 
% medium effect size only
% latex table generated in R 3.5.0 by xtable 1.8-2 package
% Wed Sep 18 10:11:09 2019
\begin{tabular}{rlrrrrrr}
 N & Method & Bias & 
 \begin{tabular}{@{}r@{}}Monte \\ Carlo SD \end{tabular} & 
 \begin{tabular}{@{}r@{}}SE \\ Estimate \end{tabular} & 
 \begin{tabular}{@{}r@{}}CI \\ Coverage \end{tabular} 
 %Monte Carlo SD & SE estimate & CI Coverage 
 & RMSE Inflation & $\frac{\|V_{\text{true}} - \expected{\hat{V}}\|}{\|V_{\text{true}}\|}$ \\ 
  \hline
200 & LMM slopes and intercepts & -0.066 & 0.887 & 0.836 & 0.929 & 1.000 & 0.046 \\ 
 & GEE Unstructured & -0.127 & 0.900 & 0.789 & 0.906 & 1.023 & 0.257 \\ 
 & LMM intercepts only & -0.401 & 0.950 & 0.891 & 0.900 & 1.161 & 0.660 \\ 
 & GEE Exchangeable & -0.402 & 0.951 & 0.890 & 0.900 & 1.161 & 0.663 \\ 
 & GEE Independence & -0.559 & 1.047 & 0.983 & 0.881 & 1.335 & 0.904 \\ 
  1000 & LMM slopes and intercepts & -0.103 & 0.389 & 0.381 & 0.939 & 1.000 & 0.044 \\ 
 & GEE Unstructured & -0.158 & 0.396 & 0.366 & 0.902 & 1.059 & 0.246 \\ 
 & LMM intercepts only & -0.445 & 0.419 & 0.408 & 0.796 & 1.519 & 0.659 \\ 
 & GEE Exchangeable & -0.445 & 0.419 & 0.408 & 0.795 & 1.519 & 0.661 \\ 
 & GEE Independence & -0.612 & 0.464 & 0.451 & 0.699 & 1.907 & 0.904 \\ 
  5000 & LMM slopes and intercepts & -0.080 & 0.177 & 0.171 & 0.921 & 1.000 & 0.042 \\ 
 & GEE Unstructured & -0.138 & 0.178 & 0.165 & 0.848 & 1.157 & 0.243 \\ 
 & LMM intercepts only & -0.437 & 0.187 & 0.184 & 0.349 & 2.448 & 0.658 \\ 
 & GEE Exchangeable & -0.437 & 0.187 & 0.184 & 0.350 & 2.448 & 0.660 \\ 
 & GEE Independence & -0.614 & 0.204 & 0.203 & 0.134 & 3.332 & 0.904 \\ 
   \hline
\end{tabular}
}
\end{table}

\clearpage
%\end{document}

\bibliographystyle{biorefs}
\bibliography{refs}

\begin{thebibliography}{99}

\bibitem[Almirall \emph{and others}(2016)Almirall, DiStefano, Chang, Shire,
  Kaiser, Lu, Nahum-Shani, Landa, Mathy and Kasari]{almirall2016longitudinal}
\textsc{Almirall, D., DiStefano, C., Chang, Y., Shire, S., Kaiser, A., Lu, X.,
  Nahum-Shani, I., Landa, R., Mathy, P. and Kasari, C.} (2016).
\newblock Longitudinal effects of adaptive interventions with a
  speech-generating device in minimally verbal children with {ASD}.
\newblock {\em Journal of Clinical Child and Adolescent
  Psychology\/}~\textbf{45}, 442--456.

\bibitem[Almirall \emph{and others}(2014)Almirall, Nahum-Shani, Sherwood and
  Murphy]{almirall_IntroToSMART:2014}
\textsc{Almirall, D., Nahum-Shani, I., Sherwood, N. and Murphy, S.~A.} (2014).
\newblock Introduction to {SMART} designs for the development of adaptive
  interventions: with application to weight loss research.
\newblock {\em Translational Behavioral Medicine\/}~\textbf{4}, 260--274.

\bibitem[August \emph{and others}(2016)August, Piehler and
  Bloomquist]{august2014being}
\textsc{August, G.~J., Piehler, T.~F. and Bloomquist, M.~L.} (2016).
\newblock Being {``SMART''} about adolescent conduct problems prevention:
  executing a {SMART} pilot study in a juvenile diversion agency.
\newblock {\em Journal of Clinical Child and Adolescent
  Psychology\/}~\textbf{45}, 495--509.

\bibitem[Bates \emph{and others}(2015)Bates, M{\"a}chler, Bolker and
  Walker]{lme4-bates}
\textsc{Bates, D., M{\"a}chler, M., Bolker, B. and Walker, S.} (2015).
\newblock Fitting linear mixed-effects models using {lme4}.
\newblock {\em Journal of Statistical Software\/}~\textbf{67}, 1--48.

\bibitem[Brumback(2009)Brumback]{brumback2009note}
\textsc{Brumback, B.~A.} (2009).
\newblock A note on using the estimated versus the known propensity score to
  estimate the average treatment effect.
\newblock {\em Statistics and Probability Letters\/}~\textbf{79}, 537--542.

\bibitem[Carlin and Louis(2000)Carlin and Louis]{carlin2000empirical}
\textsc{Carlin, B.~P. and Louis, T.~A.} (2000).
\newblock Empirical {B}ayes: Past, present and future.
\newblock {\em Journal of the American Statistical Association\/}~\textbf{95},
  1286--1289.

\bibitem[Chakraborty and Moodie(2013)Chakraborty and
  Moodie]{chakraborty-moodie_DTR-Book:2013}
\textsc{Chakraborty, B. and Moodie, E.~E.} (2013).
\newblock {\em Statistical methods for dynamic treatment regimes\/}. Springer.

\bibitem[Dawson and Lavori(2008)Dawson and Lavori]{dawson2008sequential}
\textsc{Dawson, R. and Lavori, P.~W.} (2008).
\newblock Sequential causal inference: Application to randomized trials of
  adaptive treatment strategies.
\newblock {\em Statistics in medicine\/}~\textbf{27}, 1626--1645.

\bibitem[Diggle \emph{and others}(2002)Diggle, Heagerty, Liang and
  Zeger]{diggle2002analysis}
\textsc{Diggle, P., Heagerty, P., Liang, K. and Zeger, S.} (2002).
\newblock {\em Analysis of longitudinal data\/}. Oxford University Press.

\bibitem[Dziak \emph{and others}(2019)Dziak, Yap, Almirall, McKay, Lynch and
  Nahum-Shani]{dziak-binary-2019}
\textsc{Dziak, J.~J., Yap, J. R.~T., Almirall, D., McKay, J.~R., Lynch, K.~G.
  and Nahum-Shani, I.} (2019).
\newblock A data analysis method for using longitudinal binary outcome data
  from a smart to compare adaptive interventions.
\newblock {\em Multivariate Behavioral Research\/}~\textbf{54}, 613--636.

\bibitem[Fitzmaurice \emph{and others}(2012)Fitzmaurice, Laird and
  Ware]{fitzmaurice2012applied}
\textsc{Fitzmaurice, G.~M., Laird, N.~M. and Ware, J.~H.} (2012).
\newblock {\em Applied longitudinal analysis\/}, Volume 998. John Wiley and
  Sons.

\bibitem[Gibbons \emph{and others}(2010)Gibbons, Hedeker and
  DuToit]{gibbons2010advances}
\textsc{Gibbons, R.~D., Hedeker, D. and DuToit, D.} (2010).
\newblock Advances in analysis of longitudinal data.
\newblock {\em Annual review of clinical psychology\/}~\textbf{6}, 79--107.

\bibitem[Goldstein(2011)Goldstein]{goldstein2011multilevel}
\textsc{Goldstein, H.} (2011).
\newblock {\em Multilevel statistical models\/}, Volume 922. John Wiley and
  Sons.

\bibitem[Gunlicks-Stoessel \emph{and others}(2016)Gunlicks-Stoessel, Mufson,
  Westervelt, Almirall and Murphy]{gunlicks2015pilot}
\textsc{Gunlicks-Stoessel, M., Mufson, L., Westervelt, A., Almirall, D. and
  Murphy, S.~A.} (2016).
\newblock A pilot {SMART} for developing an adaptive treatment strategy for
  adolescent depression.
\newblock {\em Journal of Clinical Child and Adolescent
  Psychology\/}~\textbf{45}, 480--494.

\bibitem[Hedeker and Gibbons(2006)Hedeker and Gibbons]{hedeker2006longitudinal}
\textsc{Hedeker, D. and Gibbons, R.~D.} (2006).
\newblock {\em Longitudinal data analysis\/}, Volume 451. John Wiley and Sons.

\bibitem[Hern\'{a}n \emph{and others}(2006)Hern\'{a}n, Lanoy, Costagliola and
  Robins]{hernan/optimaldtr-via-iptw:2006}
\textsc{Hern\'{a}n, M.A., Lanoy, E., Costagliola, D. and Robins, J.~M.} (2006).
\newblock Comparison of dynamic treatment regimes via inverse probability
  weighting.
\newblock {\em Basic and Clinical Pharmacology and Toxicology\/}~\textbf{98},
  237--242.

\bibitem[Hirano \emph{and others}(2003)Hirano, Imbens and
  Ridder]{hirano2003efficient}
\textsc{Hirano, K., Imbens, G.~W. and Ridder, G.} (2003).
\newblock Efficient estimation of average treatment effects using the estimated
  propensity score.
\newblock {\em Econometrica\/}~\textbf{71}, 1161--1189.

\bibitem[Kasari \emph{and others}(2014)Kasari, Kaiser, Goods, Nietfeld, Mathy,
  Landa, Murphy and Almirall]{kasari2014communication}
\textsc{Kasari, C., Kaiser, A., Goods, K., Nietfeld, J., Mathy, P., Landa, R.,
  Murphy, S.~A. and Almirall, D.} (2014).
\newblock Communication interventions for minimally verbal children with
  autism: a sequential multiple assignment randomized trial.
\newblock {\em Journal of the American Academy of Child and Adolescent
  Psychiatry\/}~\textbf{53}, 635--646.

\bibitem[Kidwell(2014)Kidwell]{kidwell-SMARTcancerResearch:2014}
\textsc{Kidwell, K.~M.} (2014).
\newblock {SMART} designs in cancer research: past, present, and future.
\newblock {\em Clinical Trials\/}~\textbf{11}, 445--456.

\bibitem[Kosorok and Moodie(2015)Kosorok and Moodie]{kosorok2015adaptive}
\textsc{Kosorok, M.~R. and Moodie, E. E.~M.} (2015).
\newblock {\em Adaptive treatment strategies in practice: planning trials and
  analyzing data for personalized medicine\/}, Volume~21. SIAM.

\bibitem[Lavori and Dawson(2000)Lavori and
  Dawson]{lavori-dawson_biasedRandomize:2000}
\textsc{Lavori, P. and Dawson, D.A.} (2000).
\newblock A design for testing clinical strategies: biased individually
  tailored within-subject randomization.
\newblock {\em Journal of the Royal Statistical Society, Series
  A\/}~\textbf{163}, 29--38.

\bibitem[Lavori and Dawson(2014)Lavori and Dawson]{lavori2014introduction}
\textsc{Lavori, P.~W. and Dawson, R.} (2014).
\newblock Introduction to dynamic treatment strategies and sequential multiple
  assignment randomization.
\newblock {\em Clinical Trials\/}~\textbf{11}, 393--399.

\bibitem[Lei \emph{and others}(2012)Lei, Nahum-Shani, Lynch, Oslin and
  Murphy]{lei_SMART:2012}
\textsc{Lei, H., Nahum-Shani, I., Lynch, K., Oslin, D. and Murphy, S.~A.}
  (2012).
\newblock A {SMART} design for building individualized treatment sequences.
\newblock {\em Annual Review of Clinical Psychology\/}~\textbf{8}, 21--48.

\bibitem[Li and Murphy(2011)Li and Murphy]{Zhiguo-survival}
\textsc{Li, Z. and Murphy, S.~A.} (2011).
\newblock Sample size formulae for two-stage randomized trials with survival
  outcomes.
\newblock {\em Biometrika\/}~\textbf{98}, 503--518.

\bibitem[Liang and Zeger(1986)Liang and Zeger]{liang1986longitudinal}
\textsc{Liang, K. and Zeger, S.~L.} (1986).
\newblock Longitudinal data analysis using generalized linear models.
\newblock {\em Biometrika\/}~\textbf{73}, 13--22.

\bibitem[Lu \emph{and others}(2016)Lu, Nahum-Shani, Kasari, Lynch, Oslin,
  Pelham, Fabiano and Almirall]{lu:2016}
\textsc{Lu, X., Nahum-Shani, I., Kasari, C., Lynch, K.~G., Oslin, D.~W.,
  Pelham, W.~E., Fabiano, G. and Almirall, D.} (2016).
\newblock Comparing dynamic treatment regimes using repeated-measures outcomes:
  modeling considerations in {SMART} studies.
\newblock {\em Statistics in Medicine\/}~\textbf{35}, 1595--1615.

\bibitem[Molenberghs and Kenward(2007)Molenberghs and
  Kenward]{molenberghs2007missing}
\textsc{Molenberghs, G. and Kenward, M.} (2007).
\newblock {\em Missing data in clinical studies\/}, Volume~61. John Wiley and
  Sons.

\bibitem[Murphy(2005)Murphy]{murphy_SMARTsim:2005}
\textsc{Murphy, S.~A.} (2005).
\newblock An experimental design for the development of adaptive treatment
  strategies.
\newblock {\em Statistics in Medicine\/}~\textbf{24}, 1455--1481.

\bibitem[Murphy \emph{and others}(2007)Murphy, Lynch, Oslin, McKay and
  Tenhave]{murphy-lynch-DAD:2007}
\textsc{Murphy, S.~A., Lynch, K.~G., Oslin, D.~W., McKay, J.R. and Tenhave,
  T.~R.} (2007).
\newblock Developing adaptive treatment strategies in substance abuse research.
\newblock {\em Drug and Alcohol Dependence\/}~\textbf{88}, S24--S30.

\bibitem[Murphy \emph{and others}(2001)Murphy, van~der Laan, Robins and
  {Conduct Problems Prevention Research Group}]{murphy/laan/robins/cpprg:01}
\textsc{Murphy, S.~A., van~der Laan, M.~J., Robins, J.~M. and {Conduct Problems
  Prevention Research Group}}. (2001).
\newblock Marginal mean models for dynamic regimes.
\newblock {\em Journal of the American Statistical Association\/}~\textbf{96},
  1410--1423.

\bibitem[Naar-King \emph{and others}(2016)Naar-King, Ellis, Idalski~Carcone,
  Templin, Jacques-Tiura, Brogan~Hartlieb, Cunningham and
  Jen]{naar2015sequential}
\textsc{Naar-King, S., Ellis, D.~A., Idalski~Carcone, A., Templin, T.,
  Jacques-Tiura, A.~J., Brogan~Hartlieb, K., Cunningham, P. and Jen, K-L.~C.}
  (2016).
\newblock Sequential multiple assignment randomized trial ({SMART}) to
  construct weight loss interventions for {A}frican {A}merican adolescents.
\newblock {\em Journal of Clinical Child and Adolescent
  Psychology\/}~\textbf{45}, 428--441.

\bibitem[Nahum-Shani \emph{and others}(2012)Nahum-Shani, Qian, Almirall,
  Pelham, Gnagy, Fabiano, Waxmonsky, Yu and
  Murphy]{nahum_SMARTprimaryPsychMeth:2012}
\textsc{Nahum-Shani, I., Qian, M., Almirall, D., Pelham, W.E., Gnagy, B.,
  Fabiano, G., Waxmonsky, J., Yu, J. and Murphy, S.~A.} (2012).
\newblock Experimental design and primary data analysis methods for comparing
  adaptive interventions.
\newblock {\em Psychological Methods\/}~\textbf{17}, 457--477.

\bibitem[NeCamp \emph{and others}(2017)NeCamp, Kilbourne and
  Almirall]{necamp2017comparing-DTRs-cluster-SMART}
\textsc{NeCamp, T., Kilbourne, A. and Almirall, D.} (2017).
\newblock Comparing cluster-level dynamic treatment regimens using sequential,
  multiple assignment, randomized trials: regression estimation and sample size
  considerations.
\newblock {\em Statistical methods in medical research\/}~\textbf{26},
  1572--1589.

\bibitem[Orellana \emph{and others}(2010)Orellana, Rotnitzky and
  Robins]{orellana-rot-robins_dynamicMSM_partI:2010}
\textsc{Orellana, L., Rotnitzky, A. and Robins, J.~M.} (2010).
\newblock Dynamic regime marginal structural mean models for estimating optimal
  dynamic treatment regimes, {Part} {I}: main content.
\newblock {\em International Journal of Biostatistics\/}~\textbf{6}, Article 8.

\bibitem[Raudenbush and Bryk(2002)Raudenbush and
  Bryk]{raudenbush2002hierarchical}
\textsc{Raudenbush, S.~W. and Bryk, A.~S.} (2002).
\newblock {\em Hierarchical linear models: applications and data analysis
  methods\/}, Volume~1. Sage.

\bibitem[Robinson(1991)Robinson]{robinson1991}
\textsc{Robinson, G.~K.} (1991).
\newblock That {BLUP} is a good thing: The estimation of random effects.
\newblock {\em Statistical Science\/}~\textbf{6}, 15--32.

\bibitem[Searle \emph{and others}(2006)Searle, Casella and
  McCulloch]{searlevariance}
\textsc{Searle, S.~R., Casella, G. and McCulloch, C.~E}. (2006).
\newblock {\em Variance Components\/}. Wiley.

\bibitem[Seewald \emph{and others}(2018)Seewald, Kidwell, {Nahum-Shani}, {Wu},
  {McKay} and {Almirall}]{seewald2018-repeatedmeasures-power-compareDTRs}
\textsc{Seewald, N.~J., Kidwell, K.~M., {Nahum-Shani}, I., {Wu}, T., {McKay},
  J.~R. and {Almirall}, D.} (2018).
\newblock {Sample size considerations for comparing dynamic treatment regimens
  in a sequential multiple-assignment randomized trial with a continuous
  longitudinal outcome}.
\newblock {\em arXiv e-prints\/}, arXiv:1810.13094.

\bibitem[Skrondal and Rabe-Hesketh(2009)Skrondal and
  Rabe-Hesketh]{skrondal2009prediction}
\textsc{Skrondal, A. and Rabe-Hesketh, S.} (2009).
\newblock Prediction in multilevel generalized linear models.
\newblock {\em Journal of the Royal Statistical Society---Series
  A\/}~\textbf{172}, 659--687.

\bibitem[Snijders and Bosker(2012)Snijders and Bosker]{snijdersmultilevel}
\textsc{Snijders, T. A.~B. and Bosker, R.~J.} (2012).
\newblock {\em Multilevel analysis: an introduction to basic and advanced
  multilevel modeling\/}. Sage.

\bibitem[Verbeke and Molenberghs(2009)Verbeke and
  Molenberghs]{verbeke2009linear}
\textsc{Verbeke, G. and Molenberghs, G.} (2009).
\newblock {\em Linear mixed models for longitudinal data\/}. Springer Science
  and Business Media.

\end{thebibliography}

\end{document}